\documentclass[twocolumn,preprintnumbers,amsmath,amssymb,superscriptaddress,10pt]{revtex4}
\usepackage{amssymb}
\usepackage{graphicx,color}% Include figure files
\usepackage{bm}% bold math

\begin{document}
\title{Persistent nonequilibrium effects in generalized
Langevin dynamics of nonrelativistic and relativistic particles}

\author{Weiguo Chen}
\affiliation{Department of Physics, Tsinghua University and Collaborative Innovation\\ Center of Quantum Matter, Beijing 100084, China}

\author{Carsten Greiner}
\affiliation{Institut f$\ddot{u}$r Theoretische Physik, Johann Wolfgang Goethe-Universit$\ddot{a}$t Frankfurt, Max-von-Laue-Strasse 1, 60438 Frankfurt am Main, Germany}

\author{Zhe Xu \footnote{xuzhe@mail.tsinghua.edu.cn}}
%\author{Zhe Xu}
\affiliation{Department of Physics, Tsinghua University and Collaborative Innovation\\ Center of Quantum Matter, Beijing 100084, China}
%\email{xuzhe@mail.tsinghua.edu.cn}

%\date{\today}

\begin{abstract}
Persistent nonequilibrium effects such as the memory of the initial state, the ballistic diffusion,
and the break of the equipartition theorem and the ergodicity in Brownian motions are 
investigated by analytically solving the generalized Langevin equation of nonrelativistic
Brownian particles with colored noise. These effects can also be observed in the Brownian
motion of relativistic particles by numerically solving the generalized Langevin equation for
specially chosen memory kernels. Our analyses give rise to think about
the possible anomalous motion of heavy quarks in relativistic heavy-ion collisions.
\end{abstract}
\maketitle

\section{Introduction}\label{introduction}
The Brownian motion is a famous stochastic process exhibiting the relationship
between the fluctuation and the dissipation in statistical physics. Mathematically, it is described  
by the Langevin equation. The common knowledge from text books by solving the Langevin
equation is that the Brownian motion is a random motion of the Brownian particle in 
a fluid or gas. In the long-time limit and with the ensemble average, the Brownian particle will
reach the thermal equilibrium with the matter where it is suspended and diffuse linearly with
time. On the other hand, for specific interactions between the Brownian particle and
the particles which the matter is made of, the fluctuation as well as the dissipation correlate
with their former values in the motion. The current motion is determined by the motion in 
the past by means of the memory kernel in the generalized Langevin equation.
This memory property can give rise to a different diffusion from the linear one, called
the anomalous diffusion
\cite{diffusion,diffusion1,diffusion2,wang2020fractional,PhysRevE.104.024115,wang2022anomalous,diffusion&ergodicity, 2006cond.mat..5093B, PhysRevLett.105.100602, PhysRevE.74.041125, KhinchinTheoremAnomalousDiffusion},
and break the equipartition theorem and the ergodicity \cite{ergodicity, conditionErgodicity}.
Previous research motivates us to make a systematical study of classifying memory
kernels according to the resulting different behaviors in Brownian motions.
In this way we try to find out the general behavior of the thermal equilibrium, 
diffusion, and ergodicity of the Brownian particle for any kind of memory kernel.

The motion of heavy quarks in the quark-gluon plasma produced in relativistic heavy-ion
collisions has been considered a Brownian motion in the QCD matter and described by
the relativistic form of the Langevin equation \cite{Li:2020umn, Liu:2020cpj, Ruggieri:2022kxv}. 
However, this treatment cannot simultaneously explain the experimental data of the energy loss
and the collective flow of hadronic particles stemming from heavy quarks
\cite{Moore:2004tg, Li:2019lex, PhysRevC.79.054907, Das:2015ana, Sun:2019fud, Dong:2019unq, Cao:2013ita, Cao:2014dja, Cao:2015hia, Li:2020umn, Das:2013kea, vanHees:2005wb}.
Could an anomalous motion of heavy quarks explain the data? To answer this question,
one has to first verify that the persistent nonequilibrium effects that can occur in the motion of
nonrelativistic Brownian particles can also occur in the motion of relativistic Brownian particles.
This is another goal of this paper. The verification is not trivial, since the relativistic Langevin
equation is not a linear equation and cannot be solved analytically.

The paper is organized as follows. In Sec. \ref{nonrel}, we analytically solve the generalized
Langevin equation of nonrelativistic Brownian particles by employing the Laplace
technique. The memory kernels are classified into four categories. All the behaviors are
normal in the first category. The diffusion is normal. The equipartition theorem and the 
ergodicity hold. In the second category, the memory effect occurs, the diffusion is
ballistic, and the equipartition theorem and ergodicity are broken. In the third category,
besides the memory effect and the break of the ergodicity,  an oscillating behavior appears,
which brings the Brownian particle out of and back to the equilibrium periodically.
In the fourth category, subdifussion and superdiffusion are discussed.
We give examples of the memory kernels and present the analytical results
of the averaged kinetic energy, displacement squared, and the velocity correlation function of
the Brownian particle. In Sec. \ref{relativity}, the Langevin equation of relativistic Brownian
particles is solved numerically for the memory kernels given in the previous section.
Except for the oscillation, other persistent nonequilibrium effects are also seen in relativistic
Brownian motions. We give a summary in Sec. \ref{sum} and show the details in Laplace
transformations in the Appendix.

\section{The Langevin equation of nonrelativistic Brownian particles}\label{nonrel}
\subsection{Analytical solutions}
The motion of nonrelativistic Brownian particles is described by the generalized Langevin 
equation
\begin{equation}
\label{langevin2}
m\dot{\boldsymbol{v}}(t)=-\int^{t}_0 dt' \, \Gamma(t-t') \boldsymbol{v}(t')+\boldsymbol{\xi}(t)
\end{equation}
with
\begin{equation}
\label{correl2}
 <\xi_i(t) \xi_j(t')>=\delta_{ij} A(t-t') \,.
\end{equation}
Here we set the initial time to be $0$. The generalized Langevin equation describes 
a non-Markov process. The interactions of the Brownian particle with
the molecules of the matter at earlier times also affect the current change of momentum.
This is, on the one hand, revealed by the memory kernel $\Gamma(t-t')$. On the other hand,
the noise $\boldsymbol{\xi}$ at the current time correlates with those at earlier times as
$A(t-t')$. Since the Fourier transform of $A(t-t')$ to the frequency space has a structure other
than a constant, such noise is denoted as colored noise. The fluctuation-dissipation
theorem reads
\begin{equation}
\label{dft2}
A(t-t')=k_B T \, \Gamma(t-t') \,,
\end{equation}
according to the second kind of Kubo's law~\cite{Kubo_1966}.
We note that the correlation function $A(t-t')$ [also $\Gamma(t-t')$] is such kind of function,
which Fourier transforms are non-negative  \cite{Greiner:1998vd, Xu:1999aq}.

Particularly, for $A(t-t')=\alpha \delta(t-t')$ and $\Gamma(t-t')=2\gamma \delta(t-t')$, the
generalized Langevin equation, Eq. (\ref{langevin2}), reduces to
\begin{equation}
\label{langevin1}
m\dot{\boldsymbol{v}}=-\gamma \boldsymbol{v} + \boldsymbol{\xi}\,,
\end{equation}
with
\begin{equation}
\label{correl1}
 <\xi_i(t) \xi_j(t')>=\delta_{ij} \alpha \delta(t-t') \,.
\end{equation}
This noise is called white noise. The fluctuation-dissipation theorem reduces to 
\begin{equation}
\label{dft}
\alpha=2k_BT\gamma\,.
\end{equation}

The question now is whether the Brownian particle under colored noise will relax to
the thermal equilibrium state. 
In the following, we will show that the answer to this question depends on the actual
functional form of the memory kernel $\Gamma(t-t')$. For some classes of $\Gamma(t-t')$,
the Brownian particle will not relax to the thermal equilibrium with the surrounding matter.
Moreover, even in the long-time limit the Brownian particle
still keeps the memory on its initial state, its diffusion shows an anomalous behavior, and
the ergodicity is broken.

Performing the Laplace transformation of Eq. (\ref{langevin2}), we obtain
\begin{equation}
m s \boldsymbol{v}(s)-m \boldsymbol{v}(0)= -\Gamma (s) \boldsymbol{v}(s) +
\boldsymbol{\xi}(s) \,,
\end{equation}
where $s$ is defined in the complex space. We then have
\begin{equation}
\label{laplace1}
\boldsymbol{v}(s)=\frac{\boldsymbol{v}(0)+\frac{1}{m} \boldsymbol{\xi}(s)}{s+\frac{1}{m}\Gamma (s)} \,.
\end{equation}
Defining the response function in the Laplace space as
\begin{equation}
\label{retardedGFinS}
G(s)=\frac{1}{s+\frac{1}{m}\Gamma (s)} \,,
\end{equation}
Eq. (\ref{laplace1}) is rewritten to
\begin{equation}
\label{laplace2}
\boldsymbol{v}(s)=\boldsymbol{v}(0) G(s)+\frac{1}{m}G(s) \boldsymbol{\xi}(s) \,.
\end{equation}
We then perform the inverse Laplace transformation of Eq. (\ref{laplace2}) and obtain
\begin{equation}
\label{solution2}
\boldsymbol{v}(t)=\boldsymbol{v}(0)\, G(t) +\frac{1}{m}\int ^t_0 dt' G(t-t') \boldsymbol{\xi}(t')\,.
\end{equation}
At $t=0$, we find $G(t=0)=1$. With Eq. (\ref{solution2}), we have
\begin{eqnarray}
<v^2>(t)&=&v^2(0)G^2(t)+\frac{1}{m^2}\int_0^t dt' G(t-t')  \nonumber \\
&& \times \int_0^t dt'' G(t-t'') <\boldsymbol{\xi}(t'')\cdot \boldsymbol{\xi}(t')> \nonumber \\
&=&v^2(0)G^2(t)+\frac{3k_BT}{m^2}\int_0^t dt' G(t-t')   \nonumber \\
&& \times \int_0^t dt'' G(t-t'') \Gamma(t''-t') \,.
\label{v2-1}
\end{eqnarray}
After some steps, which can be found in the Appendix, we get the final result~\cite{lapas2015non,KhinchinTheoremAnomalousDiffusion,lapas2007entropy}:
\begin{equation}
\label{v2-2}
<v^2>(t)=v^2(0) G^2(t) +\frac{3k_B T}{m}\left[1- G^2(t) \right ] \,.
\end{equation}

Before we present the results of $<v^2>(t)$  for some chosen memory kernels, we
now discuss generally the long-time behavior of $<v^2>$ and give the answer whether
the Brownian particle will relax to the thermal state. From Eq. (\ref{v2-2}), it is obvious that
if the Brownian particles are initially in the thermal state, i.e., $<v^2>(0)=3k_B T/m$, 
they always stay in the thermal state, $<v^2>(t)=3k_B T/m$, regardless of the actual form of
the memory kernel. For the case that the Brownian particles are initially out of thermal
equilibrium, they will approach the thermal state, only if $G(t\to\infty)=0$. In the following, we
examine $G(t\to\infty)$ for the nonthermal initial state of Brownian particles.

Performing the inverse Laplace transformation, we have
\begin{eqnarray}
G(t)&=&L^{-1}[G(s)] =\frac{1}{2\pi i}\int ^{\beta +i\infty}_{\beta -i\infty} ds \,G(s) e^{st} 
\nonumber \\
&=& \frac{1}{2\pi i}\int ^{\beta +i\infty}_{\beta -i\infty} ds \frac{1}{s+\frac{1}{m} \Gamma(s)}
e^{st}  \,.
\label{gt}
\end{eqnarray}
The integral area should be chosen to ensure that the integral is convergent. So, we choose
the left half of the complex plane with respect to the imaginary axis (the real part is negative).
Along the semicircle with 
the infinite radius the integral vanishes. Suppose $G(s)$ has $n$ single poles
$s_j=\sigma_j +i\omega_j$, $j=1,2,\cdots, n$. $G(s)$ can be written to
\begin{equation}
G(s)=\sum_{j=1}^n \frac{a_j}{s-s_j} \,,
\end{equation}
where 
\begin{equation}
a_j=\lim_{s\to s_j} (s-s_j) G(s) \,.
\end{equation}
%\begin{equation}
%a_j=\frac{1}{1+\frac{1}{m}\left . \frac{d\Gamma (s)}{ds} \right |_{s=s_j}} \,.
%\end{equation}
Thus, we obtain
\begin{equation}
\label{gtpole}
G(t)= \sum_{j=1}^n  a_j \, e^{s_j  t}
\end{equation}
according to the residue theorem.
The case that $G(s)$ has branch points will be discussed later in this subsection.

To discuss the general behaviors of the Brownian particle for any kind of memory kernel,
we classify memory kernels according to the position of poles and branch points of $G(s)$.
First, we consider poles only. If the poles are located in the left half of the complex plane
and not on the imaginary axis, the real part of $e^{s_j t}$ is $e^{\sigma_j t}$
with a negative $\sigma_j$. Therefore, we have $G(t\to\infty)=0$. In this case, the Brownian
particle will relax to the thermal equilibrium with the surrounding matter. For example, we
choose~\cite{KramersEscape,Schmidt:2014zpa}
\begin{equation}
\label{gm1}
\Gamma_1 (t-t') = \frac{\gamma}{2\tau} e^{-\frac{|t-t'|}{\tau}} \,.
\end{equation}
Its Laplace transform is
\begin{equation}
\label{gs1}
\Gamma_1 (s) = \frac{\gamma}{2(1+s\tau)} \,.
\end{equation}
$G_1(s)$ has two poles located in the left half of the complex plane and not on the imaginary
axis. Therefore, $G_1(t\to\infty)=0$.

Second, we consider the case that only one of the poles is located on the imaginary axis
and specially at $s=0$. In this case, it should be $\Gamma(s=0)=0$ [see 
Eq. (\ref{retardedGFinS})]. We have $G(t\to\infty)=a_1$, where
\begin{eqnarray}
\label{gamma2a1}
a_1&=&\lim_{s\to 0} s G(s)=\lim_{s\to 0} \frac{s}{s+\frac{1}{m} \Gamma(s)} \nonumber \\
&=&\frac{1}{1+\frac{1}{m} \left . \frac{d\Gamma(s)}{ds} \right |_{s=0}} \equiv \frac{1}{1+Q}
\end{eqnarray}
with
\begin{eqnarray}
Q&=&\frac{1}{m} \left . \frac{d}{ds} \int_0^\infty dt \, \Gamma(t) e^{-st} \right |_{s=0}
\nonumber \\
&=&-\frac{1}{m} \int_0^\infty dt \, \Gamma(t) t \,.
\end{eqnarray}
We see that $G(t\to\infty)$ is nonzero. The equipartition theorem is broken and the Brownian
particle will relax to a certain state, but not to the thermal equilibrium with the surrounding matter.
In addition, from Eq. (\ref{v2-2}) we see that the term $v^2(0) G^2(t)$ contributes to $<v^2>(t)$,
which indicates that in the long-time limit the Brownian particle still keeps the memory
of its initial state. This is the memory effect.

From Eq. (\ref{v2-2}), we also see that $G(t)$ is smaller than $1$, because $<v^2>$ is
always positive, also for $\boldsymbol{v}(0)=0$. If $\boldsymbol{v}(0)=0$, the kinetic energy
$m<v^2>/2$ will reach a smaller value than that from the equipartition theorem.
$G(t) <1$ also leads to $a_1 < 1$ and thus $Q>0$.
We realize that in this case negative correlations, $\Gamma(t) <0$ at some time interval,
will occur. This indicates that the mean force does not always decelerate the Brownian particle.
It will also accelerate the Brownian particle. This might be the physical reason, 
why the Brownian particle cannot approach the thermal equilibrium with the surrounding 
matter. 

We choose, for example~\cite{KramersEscape},
\begin{equation}
\label{gm2}
\Gamma_2 (t-t')= \frac{\gamma}{4\tau} \left ( 1-\frac{|t-t'|}{\tau} \right ) e^{-\frac{|t-t'|}{\tau}} \,,
\end{equation}
which falls to be negative at $t-t'=\tau$ and approaches $0$ at large $t-t'$ from
the negative side. This correlation resembles that for studying the non-Markov
dissipative evolution of the chiral fields \cite{Xu:1999aq}. The Laplace transform is
\begin{equation}
\label{gs2}
\Gamma_2 (s)=\frac{\gamma s \tau}{4(1+s \tau )^2}\,.
\end{equation}
$G_2(s)$ has one pole at $s=0$ and the other two poles in the left half of the complex plane
and not on the imaginary axis.

Third, some poles are located on the imaginary axis, but not at $s=0$. These poles
appear in pairs symmetric to $s=0$, $s_{1,2}=\pm i \omega_1$, for instance. We can 
write $G(s)$ to the form
\begin{equation}
G(s)=\sum_{j=1}^k \left ( \frac{a_j}{s-i\omega_j}+  \frac{a_j^*}{s+i\omega_j} \right )+ 
\sum_{j=2k+1}^n \frac{a_j}{s-s_j} \,.
\end{equation}
We obtain 
$G(\infty)=\lim_{t\to \infty}\sum_{j=1}^k 2[Re (a_j) \cos(\omega_j t) -Im(a_j)\sin(\omega_j t)]$.
The oscillations of $G(t)$ are not damped in the long-time limit due to the absence of the real
part of the poles. In this case, the Brownian particle will not even approach
a steady state. On the other hand, $G(t)$ oscillates across zero. When $G(t)$
is zero, the Brownian particle is at the thermal equilibrium. So, in the long-time limit,
the Brownian particle goes out of and returns to the thermal equilibrium circularly.
An example,  $\Gamma_3$, will be given later.
We note that kernels of the second and third classes were not discussed in 
Ref. \cite{book:18281},
where the Fourier transformation was used to solve the generalized Langevin equation.

Fourth, we consider the memory kernels having the following form at $s\to 0$ \cite{KhinchinTheoremAnomalousDiffusion}:
\begin{equation}
\label{gs4}
\lim_{s\to 0} \Gamma_4(s)=\gamma(s\tau)^{\lambda-1} \,,
\end{equation}
where $\lambda$ is a fractional number. With this, $G_4(s)$ has the branch points at $s=0$.
For $G(s)$ having branch points, but other than Eq. (\ref{gs4}), we postpone to
further investigations. Putting Eq. (\ref{gs4}) into Eq. (\ref{gt}), we obtain
\begin{equation}
\label{gt2}
G_4(t)=\frac{1}{2\pi i}\int d(st) \frac{(st)^{1-\lambda}}{(st)^{2-\lambda}+\frac{\gamma\tau}{m} 
\left ( \frac{t}{\tau} \right)^{2-\lambda}} e^{st} \,.
\end{equation}
In the long-time limit and $\lambda < 2$, $G_4(t)$ can be solved by using asymptotic
expansions \cite{haubold2011mittag}. For $0<\lambda < 2$, it is
\begin{equation}
\label{gt3}
G_4(t) \sim \left ( \frac{t}{\tau} \right)^{\lambda-2} \to 0 \,.
\end{equation}
The Brownian particle will relax to the thermal equilibrium with the surrounding matter and
the equipartition theorem of energy holds.
For $\lambda < 0$, we have
\begin{equation}
G_4(t) \sim \sum_n 
\left ( \frac{t}{\tau} \right )^{\frac{2-\lambda}{n}} \exp\left [ \exp \left ( \frac{2n\pi i}{2-\lambda}
\right ) \frac{t}{\tau} \right ] \,,
\end{equation}
which is complex and divergent for $t\to \infty$. Thus, this case is not physical. 
At last, for $\lambda > 2$, $(t/\tau)^{2-\lambda}$ goes to $0$ in the long-time limit and
thus, $s=0$ becomes a pole position [see Eq. (\ref{gt2})]. The persistent nonequilibrium
effects are the same as those discussed before for the class of memory kernels with a pole
at $s=0$. 

We present now the analytical results of $G(t)$ for chosen memory kernels 
$\Gamma(t-t')$. For $\Gamma_1(t-t')$ [see Eq. (\ref{gm1})], we obtain results for two cases.
If $B\equiv \sqrt{1-2\gamma \tau/m}$ is real, we have
\begin{equation}
\label{g11t}
G_{11} (t) = \left [ \cosh \left ( B\frac{t}{2\tau} \right ) +\frac{1}{B}\sinh \left ( B\frac{t}{2\tau}
\right ) \right ] e^{-\frac{t}{2\tau}} \,.
\end{equation}
If $B$ is pure imaginary, we define $B'=-iB=\sqrt{2\gamma\tau/m-1}$ and have
\begin{equation}
\label{g12t}
G_{12}(t)= \left [ \cos \left ( B'\frac{t}{2\tau} \right ) +\frac{1}{B'}\sin \left ( B'\frac{t}{2\tau}
\right ) \right ] e^{-\frac{t}{2\tau}} \,.
\end{equation}
For $\Gamma_2(t-t')$ [see Eq. (\ref{gm2})], we obtain
\begin{eqnarray}
\label{g2t}
G_2(t)&=&\frac{1}{1+C}+\left [ \frac{C}{1+C} \cos\left ( \sqrt{C} \frac{t}{\tau} \right ) \right .
\nonumber \\
&& \left .+ \frac{\sqrt{C}}{1+C} \sin\left ( \sqrt{C} \frac{t}{\tau} \right ) \right ] e^{-\frac{t}{\tau}} \,,
\end{eqnarray}
where $C=\gamma \tau/(4m)$.

For the third case, we choose
\begin{equation}
\label{gm3}
\Gamma _3 (t-t')= \gamma \frac{1+\omega^2 \tau^2}{2\tau(\omega^2\tau^2+1-\epsilon)} 
[1-\epsilon \cos(\omega |t-t'|)] e^{-\frac{|t-t'|}{\tau}} \,,
\end{equation}
which has a oscillating disturbance. The amplitude $\epsilon$ should be 
\begin{equation}
\label{epsilon_condition1}
\epsilon \le \frac{4}{W} \left [ \sqrt{(W+1)(W+4)}-W -2 \right ] 
\end{equation}
with $W=\omega^2\tau^2$,
so the Fourier transform of $\Gamma_3(t-t')$ is non-negative. The Laplace transform 
of $\Gamma_3(t-t')$ is
\begin{equation}
\label{gs3}
\Gamma_3 (s)= \gamma'
\left [ \frac{1}{1+s\tau} -\epsilon \frac{1+s\tau}{W+(1+s\tau)^2} \right ] \,,
\end{equation}
where
\begin{equation}
\gamma'=\gamma \frac{W+1}{2(W+1-\epsilon)} \,.
\end{equation}
$G_3(s)$ has four poles. Two of them are located in the left half of the complex plane,
while another two poles are on the imaginary axis, when $\gamma'$ is the solution of
the equation
\begin{eqnarray}
&&2(1-\epsilon)^2 \tau^2 \gamma'^2 +m\tau [(1-\epsilon)
(8+5W)-9W] \gamma' \nonumber \\
&&+2m^2(W+1)(W+4) =0\,.
\end{eqnarray}
The condition that the above equation has a solution is
\begin{equation}
\label{epsilon_condition2}
\epsilon \ge \frac{4}{W} \left [ \sqrt{(W+1)(W+4)}-W -2 \right ] \,.
\end{equation}
Together with condition (\ref{epsilon_condition1}), we have
\begin{equation}
\label{epsilon_condition3}
\epsilon = \frac{4}{W} \left [ \sqrt{(W+1)(W+4)}-W -2 \right ] \,.
\end{equation}
We see that $\epsilon$ is fixed for given $\omega\tau$. With this, we obtain
\begin{equation}
\label{parameter}
\frac{\gamma \tau}{m}=\frac{2(W+1-\epsilon)}{W+1} \frac{\sqrt{(W+1)(W+4)}}
{1-\epsilon} \,.
\end{equation}
The strength of the memory kernel, $\gamma$, is not a free parameter. It relates to the
frequency of the oscillation $\omega$, the decay timescale $\tau$, and the mass of the
Brownian particle $m$. This indicates that the case that poles are located on the imaginary
axis (not at the origin) hardly occurs. This is different from $\Gamma_1(t-t')$ and 
$\Gamma_2(t-t')$, in which $\gamma$ and $\tau$ are free parameters.
After a lengthy derivation, we obtain
\begin{eqnarray}
\label{g3t}
G_3(t)&=& \frac{1}{1+\frac{1}{2}\sqrt{\frac{W+1}{W+4}}}
\cos\left ( \sqrt{\frac{Z-6W-3}{6}} \frac{t}{\tau} \right )
\nonumber \\
&&+ \frac{1}{1+2\sqrt{\frac{W+4}{W+1}}} \left  [ \frac{3\sqrt{3}}{\sqrt{Z}}
\sin\left ( \frac{\sqrt{Z}}{2\sqrt{3}} \frac{t}{\tau} \right ) \right . \nonumber \\
&& \left . +\cos\left ( \frac{\sqrt{Z}}{2\sqrt{3}} \frac{t}{\tau} \right ) \right ] e^{-\frac{3t}{2\tau}}\,,
\end{eqnarray}
where $Z=8W+4\sqrt{(W+1)(W+4)}+5$.

\begin{figure}[h]
\centering
\includegraphics[width = 8cm , height =10cm]{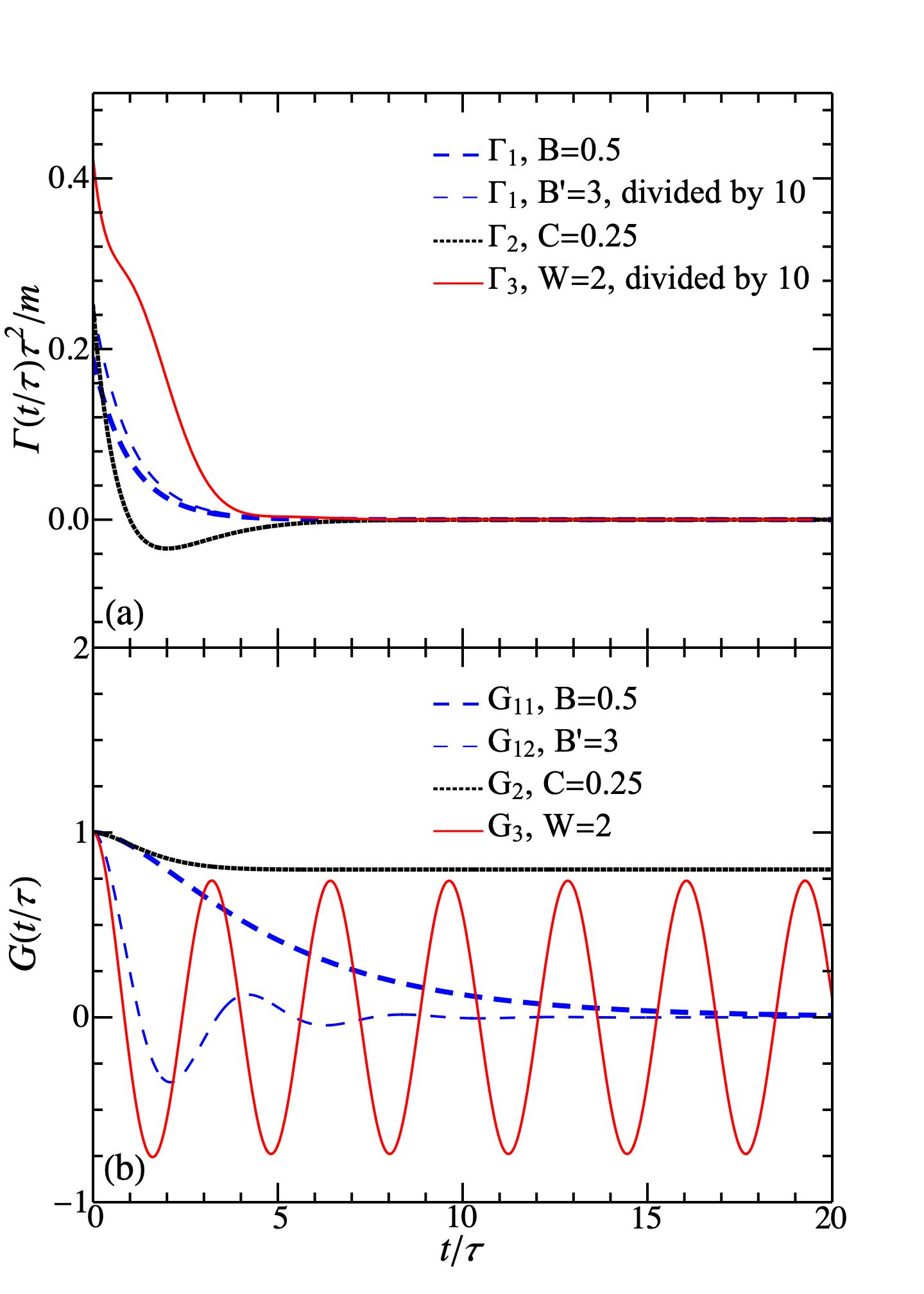}
\caption{Examples of the memory kernels (upper panel) and the response functions 
(lower panel) as functions of $t/\tau$. The memory kernels are multiplied by $\tau^2/m$.
}
\label{kernel_G}
\end{figure}
From the dependence of $G_{11}$ on $B$, $G_{12}$ on $B'$, $G_2$ on $C$, and 
$G_3$ on $W$, we realize that all $G$'s, as functions of $t/\tau$, are fixed by the only
parameter $\gamma \tau/m$. The same is also for the memory kernels $\Gamma_1$,
$\Gamma_2$, and $\Gamma_3$ multiplied by $\tau^2/m$.
Figure \ref{kernel_G} shows the chosen memory kernels (times $\tau^2/m$) and
the respective $G$'s as functions of $t/\tau$.
The curves of $\Gamma_1$ with $B'=3$ and $\Gamma_3$ with $W=2$ are divided by $10$. 
We do not give examples of memory kernels in the last category because $G(t)$ cannot be
solved analytically.

\begin{figure}[h]
\centering
\includegraphics[width = 8cm , height =7cm]{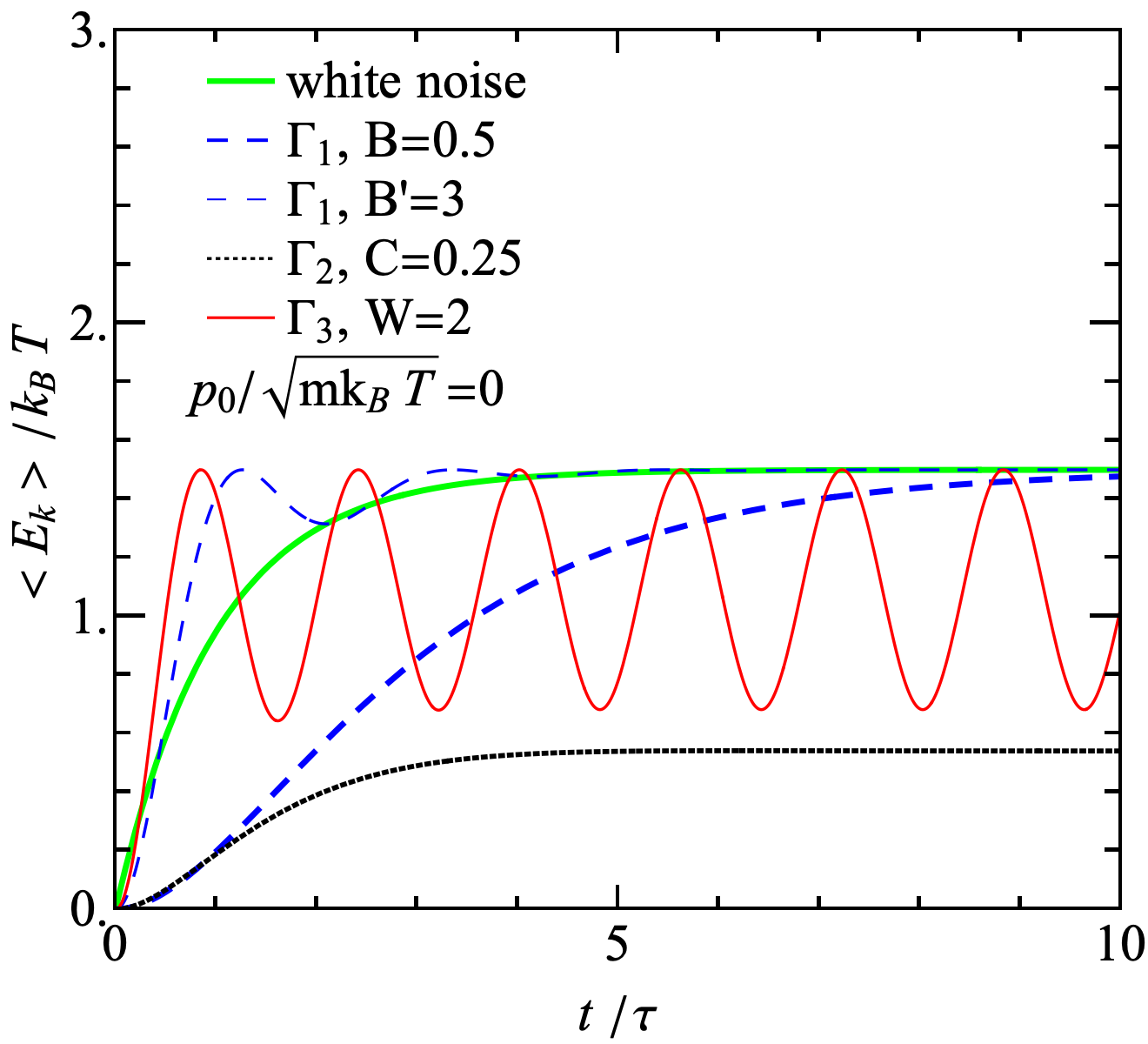}
\caption{The time evolution of the average kinetic energy of the Brownian particle under
white noise and colored noise with the chosen memory kernels. The initial
momentum of the Brownian particle is set to be zero.}
\label{kinetic_energy_p0}
\end{figure}

In Fig. \ref{kinetic_energy_p0}, we show the time evolution of the average kinetic energy
of the Brownian particle, $<E_k>=m<v^2>/2$, scaled by $k_B T$, according to Eq. (\ref{v2-2}).
The initial momentum of the Brownian particle, $p_0=mv(0)$, is set to be zero. The solid
curve depicts the result under white noise, where $G(t)=e^{-t/\tau}$.
Here $\tau$ denotes $m/\gamma$ and thus $\gamma \tau/m=1$. 
Other curves depict the results under colored noise with the memory kernels given before
and shown in Fig. \ref{kernel_G}. We see that the averaged kinetic energy of the Brownian
particle under white noise and colored noise with $\Gamma_1$ approaches
the value according to the equipartition theorem. This corresponds to the case that
the poles of the response function $G(s)$, the Laplace transform of $G(t)$, locate in the left
half of the complex plane and not on the imaginary axis. For the case with $\Gamma_2$,
where one of the poles of $G(s)$ is at the origin, $s=0$, the averaged kinetic energy is less
than the value according to the equipartition theorem. Finally, for the case with $\Gamma_3$,
where two of the poles of $G(s)$ are on the imaginary axis, the averaged kinetic energy
oscillates. At the peaks, it reaches the value according to the equipartition theorem.
The difference between the peak and the trough equals to the square of the amplitude of
the first term of $G_3(t)$ times $3/2$ [see Eqs. (\ref{g3t}) and (\ref{v2-2})] and varies between
$24/25$ for $W\to 0$ and $2/3$ for $W\to \infty$.

\begin{figure}[ht]
\centering
\includegraphics[width = 8cm , height =7cm]{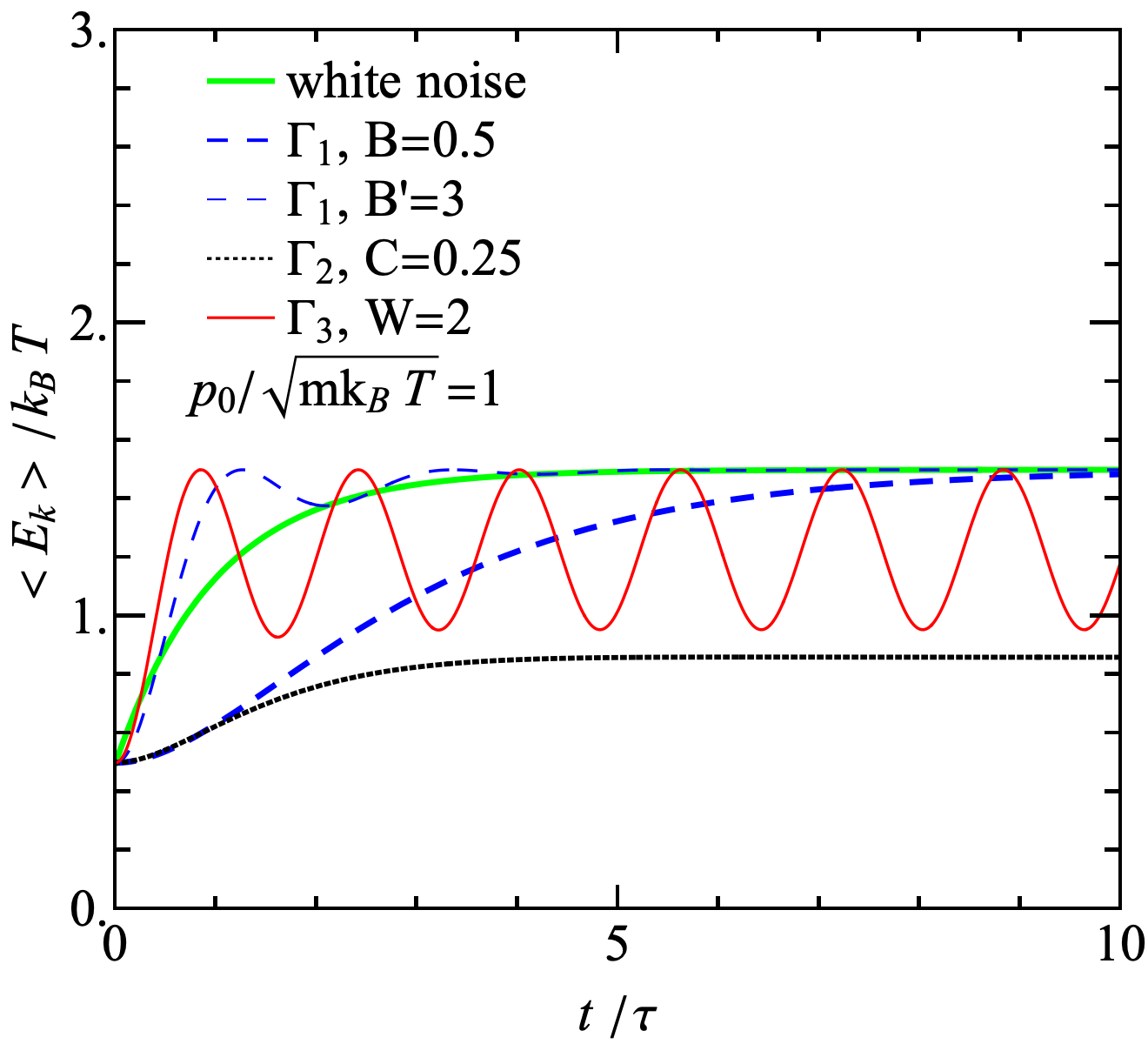}
\caption{Same as Fig. \ref{kinetic_energy_p0}. The initial momentum is set to be 
$p_0=\sqrt{mk_B T}$.}
\label{kinetic_energy_p1}
\end{figure}

By comparing with Fig. \ref{kinetic_energy_p0}, the memory effect can be observed 
in Fig. \ref{kinetic_energy_p1}, where the initial momentum is set to be $p_0=\sqrt{mk_B T}$.
Again, in the long-time limit, the averaged kinetic energy of the Brownian particle under
white noise and colored noise with $\Gamma_1$ approaches the value according
to the equipartition theorem. In these cases, no memory effect is expected. From the results
with $\Gamma_2$ and $\Gamma_3$, we do see the memory effect. The larger the initial
momentum, the larger the averaged kinetic energy of the Brownian particle with
$\Gamma_2$ in the long-time limit. In the case of $\Gamma_3$, for $p_0 < \sqrt{3mk_B T}$,
the averaged kinetic energy at long times reaches the value according to the equipartition
at the peaks of the oscillation. The difference between the peak and the trough is decreasing
to zero, when the initial momentum is increasing from zero to $\sqrt{3mk_B T}$.
For $p_0 > \sqrt{3mk_B T}$,  the averaged kinetic energy at long times reaches the value
according to the equipartition at the troughs of the oscillation. The difference between the peak
and the trough is increasing for the increasing initial momentum.

We now study the diffusion behavior of the Brownian particle with colored noise. 
Since $\boldsymbol{v}=\dot{\boldsymbol{x}}$, the Langevin equation, Eq. (\ref{langevin2}),
can be rewritten to
\begin{equation}
\label{langevin_x}
m\ddot{\boldsymbol{x}}=-\int^{t}_0 dt' \Gamma(t-t') \dot{\boldsymbol{x}}(t')+ \boldsymbol{\xi}(t)
\,.
\end{equation}
Performing the Laplace transformation to Eq. (\ref{langevin_x}) and proceeding the same steps
such as solving Eq. (\ref{langevin2}), we obtain the analytical results of 
$<(\Delta \boldsymbol{x})^2>$,
\begin{eqnarray}
\label{dx2}
<(\Delta \boldsymbol{x})^2>(t)&=&\boldsymbol{v}^2(0)H^2(t)+\frac{3k_B T}{m} 
\left [ 2 \int ^{t}_{0}dt'H(t-t') \right . \nonumber \\
&& \left. -H^2(t) \right] \,,
\end{eqnarray}
where
\begin{equation}
\label{hgs}
H(t)=L^{-1} [H(s)] \ \mbox{and} \ H(s)=\frac{1}{s} G(s)\,.
\end{equation}
Since $H(t=0)=0$, we have $\dot{H}(t)=G(t)$. For the known $G(t)$, the analytical results of
$<(\Delta \boldsymbol{x})^2>(t)$ can be easily calculated. In the following, we discuss its
long-time behavior. 
\begin{itemize}
\item[(i)] For the case that the poles of $G(s)$ are located in the left half of the complex plane
and not on the imaginary axis such as for white noise and colored noise with
$\Gamma_1$, $H(s)$ has an additional pole at $s=0$ due to Eq. (\ref{hgs}).
In the long-time limit, we have $H(t)\to G(s=0)$ and thus
$<(\Delta \boldsymbol{x})^2>(t) \to 6k_B T G(s=0) t/m$, which is proportional to $t$. This is 
the normal diffusion. For white noise, $G(s=0)=m/\gamma$, we get 
$<(\Delta \boldsymbol{x})^2>(t) \to 6k_B T t/\gamma=6Dt \sim t$, where $D$ is the diffusion
constant. For $\Gamma_1$, $G_1(s=0)=2m/\gamma$, 
we get $<(\Delta \boldsymbol{x})^2>(t) \to 12k_B T t/\gamma \sim t$. Therefore,
\begin{eqnarray}
\label{difflimit_wn}
&&\frac{m <(\Delta \boldsymbol{x})^2>|_{\mbox{wn}}}{k_B T \tau^2} \to 
6 \frac{t}{\tau} \,, \\
\label{difflimit_1} 
&&\frac{m <(\Delta \boldsymbol{x})^2>|_{\Gamma_1}}{k_B T \tau^2} \to 
12 \frac{m}{\gamma \tau} \frac{t}{\tau} \,.
\end{eqnarray}
\item[(ii)] For the case that one pole of $G(s)$ is located at $s=0$ and the other poles are
located in the left half of the complex plane and not on the imaginary axis such as for colored
noise with $\Gamma_2$, $H(s)$ has a twofold pole at $s=0$. In the long-time limit, $H(t)$
goes to the residue of $H(s)e^{st}$ at $s=0$, which is 
$d[s^2H(s)e^{st}]/ds|_{s=0} \sim sG(s)t |_{s=0}=a_1t$ [see Eq. (\ref{gamma2a1})].
This leads to 
$<(\Delta \boldsymbol{x})^2>(t) \to [v^2(0)a_1^2+3k_BTa_1(1-a_1)/m]t^2 \sim t^2$.
The diffusion with a parabolic dependence of $<(\Delta \boldsymbol{x})^2>$ on $t$ is
an anomalous diffusion and called the ballistic diffusion.
For $\Gamma_2$, we obtain
$<(\Delta \boldsymbol{x})^2>(t) \to [\boldsymbol{v}^2(0)+3k_B T C/m] t^2/(1+C)^2$
and
\begin{equation}
\label{difflimit_2}
\frac{m <(\Delta \boldsymbol{x})^2>|_{\Gamma_2}}{k_B T \tau^2} \to
 \frac{m\boldsymbol{v}^2(0)/(k_B T)+3C}{(1+C)^2} \left ( \frac{t}{\tau} \right )^2\,. 
\end{equation}
\item[(iii)] For the case that some pols of $G(s)$ are located on the imaginary axis but not
at $s=0$ and the other poles are located in the left half of the complex plane such as for 
colored noise with $\Gamma_3$, the poles on the imaginary axis lead to the oscillation of
$H(t)$ with constant amplitudes. The long-time behavior is dominated by the pole of $H(s)$
at $s=0$ like in case (i). For $\Gamma_3$, we get $G_3(s=0)=2m/\gamma$,
$<(\Delta \boldsymbol{x})^2>(t) \to 12k_B T t/\gamma \sim t$, and
\begin{equation}
\label{difflimit_3}
\frac{m <(\Delta \boldsymbol{x})^2>|_{\Gamma_3}}{k_B T \tau^2} \to 
12 \frac{m}{\gamma \tau} \frac{t}{\tau} \,. 
\end{equation}
This is again the normal diffusion.
\item[(iv)] For the case that the memory kernels $\Gamma(s)$ having branch points such as 
Eq. (\ref{gs4}), $G_4(t) \sim (t/\tau)^{\lambda-2}$ for $0< \lambda <2$ in the long-time limit
[see Eq. (\ref{gt3})]. $H_4(t)$ is obtained by the integral $H_4(t)=\int G_4(t) dt$.
Putting $H_4(t)$ into Eq. (\ref{dx2}), we have 
$<(\Delta \boldsymbol{x})^2>(t)  \sim t^\lambda$. For $0<\lambda <1$, the diffusion is called
the subdiffusion, while for $1<\lambda <2$ it is called the superdiffusion \cite{diffusion,PhysRevLett.105.100602,KhinchinTheoremAnomalousDiffusion,PhysRevE.72.067701,lapas2015non,lapas2007entropy,porra1996generalized,li2003anomalous}.
\end{itemize}

\begin{figure}[b]
\centering
\includegraphics[width = 8cm , height =7cm]{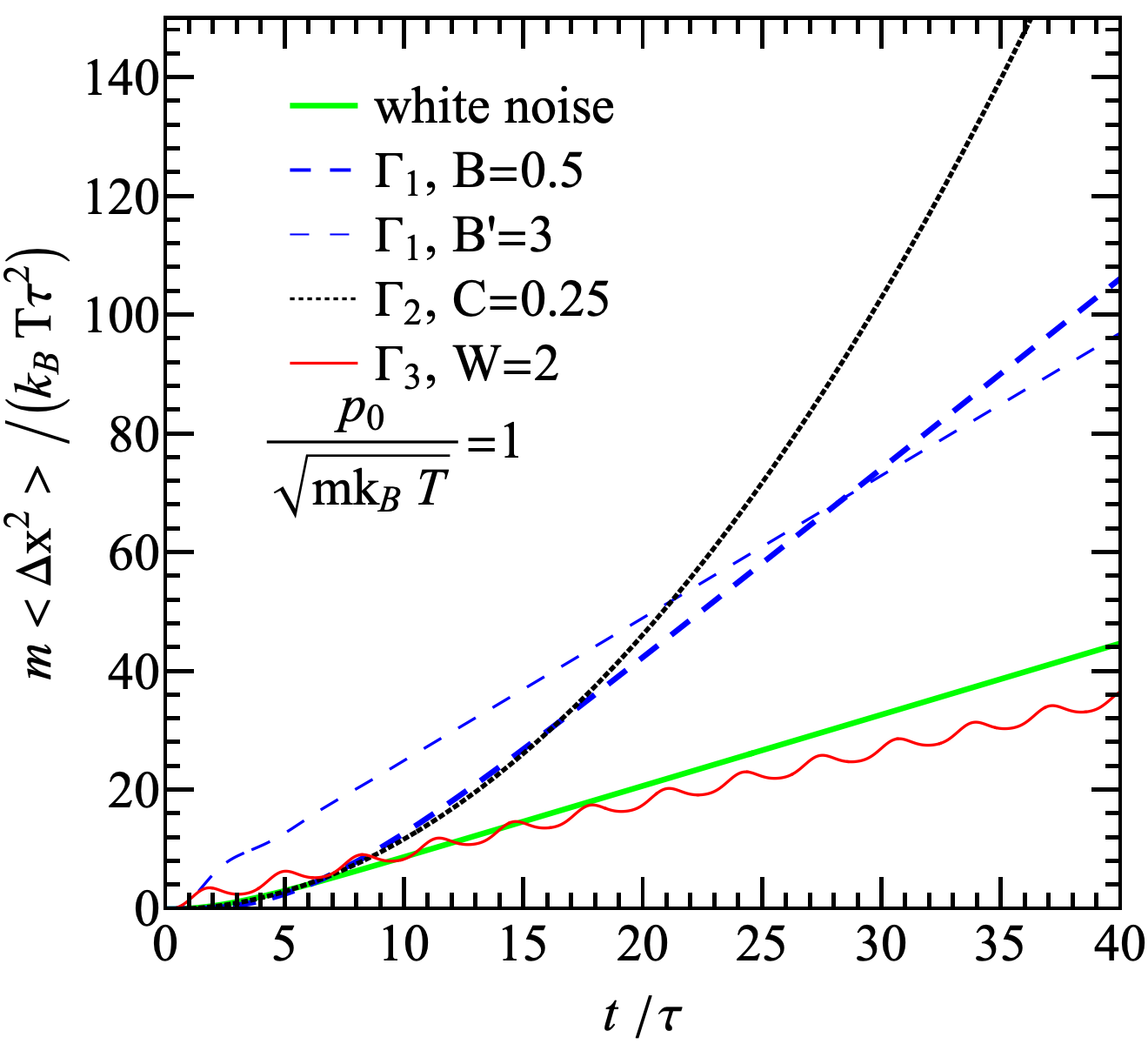}
\caption{The time evolution of the average displacement squared of the Brownian particle under
white noise and colored noise with various memory kernels. The initial
momentum of the Brownian particle is set to be $p_0=\sqrt{mk_B T}$. 
The results with white noise, $\Gamma_1$ ($B=0.5$), and $\Gamma_2$ are divided
by $10$.}
\label{x2}
\end{figure}

Figure \ref{x2} shows the time evolution of the average displacement squared of the Brownian
particle scaled by $k_B T\tau^2/m$.
The results with white noise, $\Gamma_1$ ($B=0.5$), and $\Gamma_2$ are divided
by $10$. We see that the average displacement squared of the Brownian particle under
white noise and colored noise with $\Gamma_1$ is linear in time in the long-time limit,
whereas it is parabolic in time for colored noise with $\Gamma_2$ and is oscillating along
a linear line in time for colored noise with $\Gamma_3$.

Finally, we discuss briefly the ergodicity in the Brownian motion with colored noise.
The ergodicity states that the ensemble average of a variable equals its time
average in the infinite-time limit. The necessary and sufficient condition of the ergodicity
is given by Khinchin's theorem \cite{KhinchinTheoremAnomalousDiffusion,conditionErgodicity}, 
which declares that $C_O(t)/C_O(0)$ should vanish in the limit $t \to \infty$.
$C_{O}(t)=<O(t)O(0)>-<O(t)><O(0)>$ is the autocorrelation of a stochastic variable $O(t)$.
In the following, we examine $C_{\boldsymbol{v}}(t)$ for various given memory kernels.  

The derivation of $<\boldsymbol{v}(t)\cdot \boldsymbol{v}(0)>$ from Eq. (\ref{solution2})
is similar to that of $<v^2>(t)$. The result \cite{KhinchinTheoremAnomalousDiffusion} is
\begin{equation}
C_{\boldsymbol{v}}(t)/C_{\boldsymbol{v}}(0) =G(t)\,.
\end{equation}
For white noise and colored noise with the first and fourth class of the memory kernels
such as $\Gamma_1(t-t')$ and $\lim_{s\to 0} \Gamma_4(s)=\gamma(s\tau)^{\lambda-1}$ with
$0<\lambda < 2$ , $G(t\to\infty)=0$ and we have 
$\lim_{t\to \infty} C_{\boldsymbol{v}}(t)/C_{\boldsymbol{v}}(0)=0$.
The ergodicity holds.

For colored noise with other memory kernels such as $\Gamma_2$ and $\Gamma_3$,
the limits of $G_2(t)$ and $G_3(t)$ [see Eqs. (\ref{g2t}) and (\ref{g3t})] do not go to zero for
$t\to \infty$. Thus, $\lim_{t\to \infty} C_{\boldsymbol{v}}(t)$ does not go to zero either.
We see that velocities are always correlated, although the memory kernels have finite widths.
In these two cases, the ergodicity is broken. 
In general, the ergodicity is broken when $G(s)$ has poles on the imaginal axis.
This result does not depend on the initial state, which is different from the equilibration. 
As we have noticed, the equipartition theorem is broken for the cases with $\Gamma_2$ and
$\Gamma_3$, if the initial Brownian particles are out of thermal equilibrium \cite{lapas2007entropy}.

\subsection{Numerical calculations}\label{numericalResults}
In this subsection, we solve the Langevin equation, Eq. (\ref{langevin2}), numerically.
Comparing to the conventional numerical method for solving ordinary differential equations,
here we have to generate the series of noise from its time correlation [see Eqs. (\ref{correl2})
and (\ref{dft2})]. The details of the generation of white and colored noise and the numerical 
method can be found in Ref. \cite{Xu:1999aq}.

We have two purposes for performing the numerical calculations. First, we test our numerical
computations by comparing the numerical results with the analytical ones to
prepare a well-tested numerical code for solving the relativistic Langevin equation,  which is
introduced and investigated in the next section. Second, we want to calculate
the momentum distribution of the Brownian particle in the long-time limit in the cases of
colored noise with $\Gamma_2$ and $\Gamma_3$, since obviously these distributions
cannot be obtained analytically.

From the previous subsection, we notice that the Langevin equation, Eq. (\ref{langevin2}),
can be nondimensionalized. Defining the dimensionless quantities 
$\tilde{\boldsymbol{p}}=m\boldsymbol{v}/\sqrt{mk_B T}$ and $\tilde{t}=t/\tau$, we obtain
\begin{equation}
\label{langevin_nond}
\frac{d\tilde{\boldsymbol{p}}}{d\tilde{t}}=-\int^{\tilde{t}}_0 d\tilde{t}' \, 
\tilde{\Gamma}(\tilde{t}-\tilde{t}') \tilde{\boldsymbol{p}}(\tilde{t}')+
\tilde{\boldsymbol{\xi}}(\tilde{t})\,,
\end{equation}
where
\begin{equation}
\tilde{\Gamma}(\tilde{t}-\tilde{t}')= \frac{\tau^2}{m} \Gamma(t-t') \,,
\end{equation}
which is dimensionless. Examples are plotted in the upper panel of Fig. \ref{kernel_G}.
The dimensionless noise is
$\tilde{\boldsymbol{\xi}}(\tilde{t})=\boldsymbol{\xi}(t) \tau/\sqrt{mk_B T}$, with
\begin{equation}
<\tilde{\xi}_i(\tilde{t}) \tilde{\xi}_j(\tilde{t}')>=\delta_{ij} \tilde{\Gamma}(\tilde{t}-\tilde{t}')\,.
\end{equation}
Remember that $\tau$ is the decay timescale of the memory kernels. 
For the given memory kernels [see Eqs. (\ref{gm1}), (\ref{gm2}), and (\ref{gm3})]
the only free parameter in Eq. (\ref{langevin_nond}) is $\gamma \tau/m$.

\begin{figure}[h]
\centering
%\subfigure[colored noise sequence.]{
%begin{minipage}[t]{0.5 \linewidth}
%\includegraphics[width = 8cm , height =7cm]{noise_OPS_cnoise.png}
\includegraphics[width=0.43\textwidth]{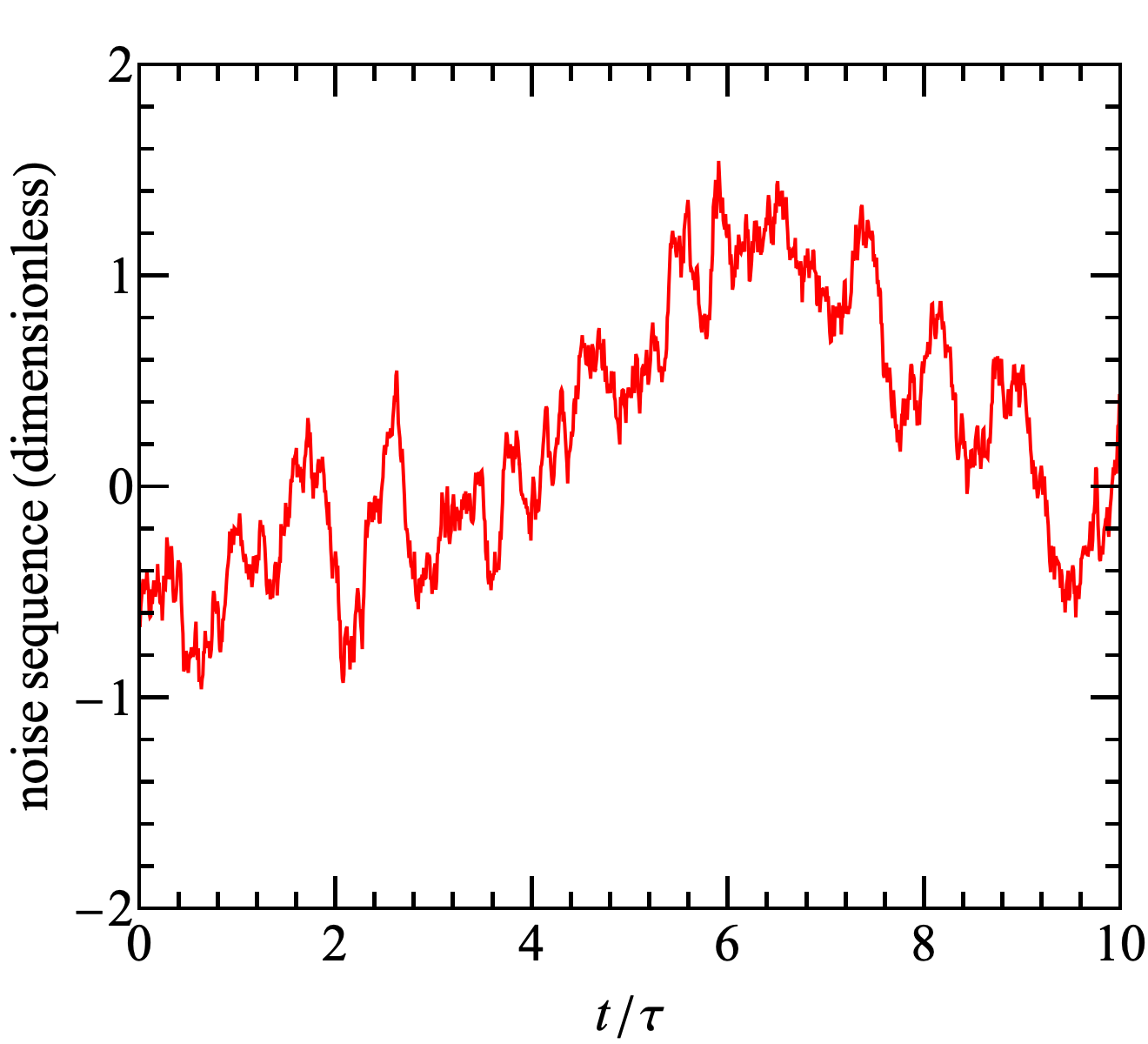}
\caption{Colored noise sequence.}
\label{noise}
%\end{minipage}
\end{figure}

\begin{figure}[h]
%\subfigure[noise correlation function.]{
%\begin{minipage}[t]{0.5 \linewidth}
\centering
\includegraphics[width = 8cm , height =7cm]{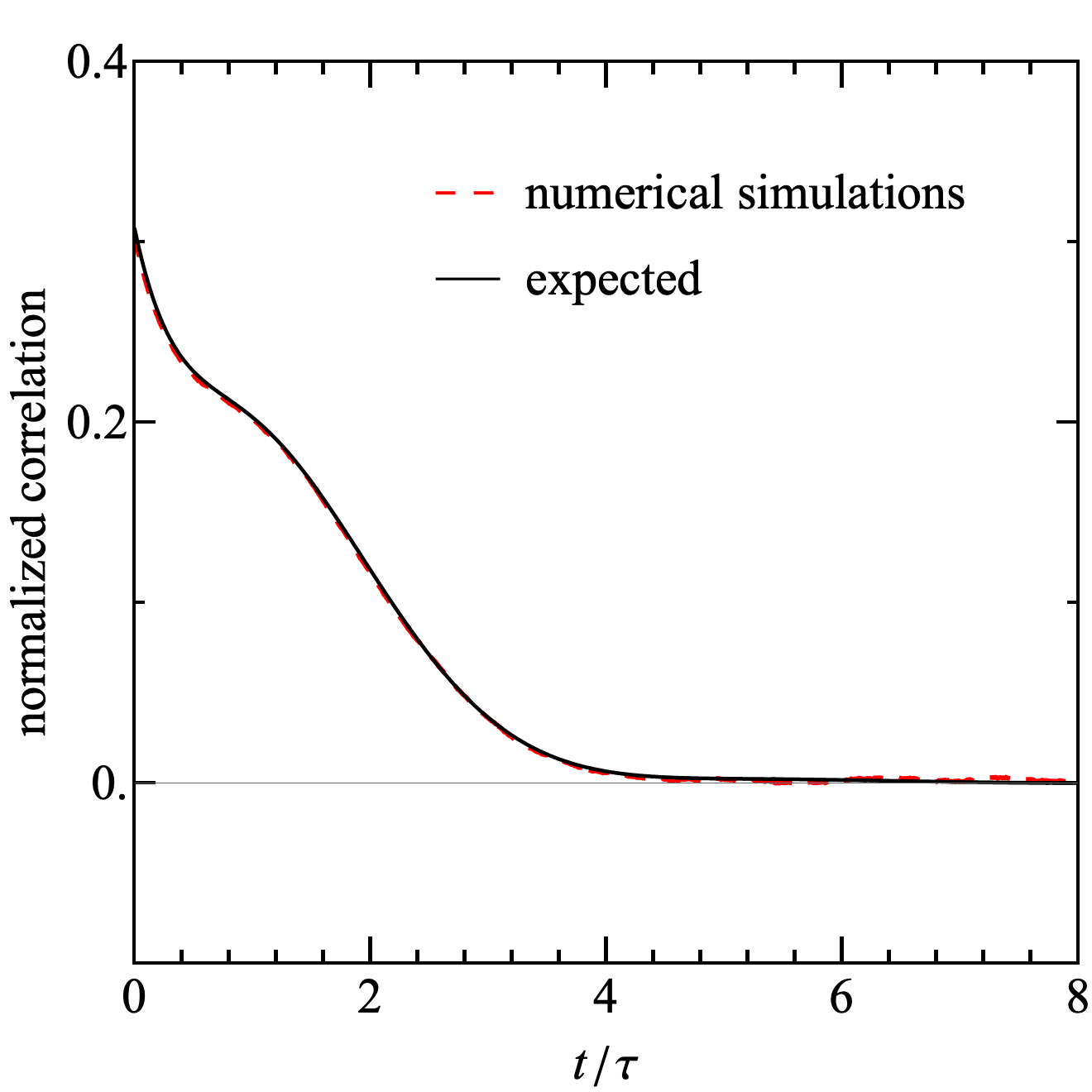}
%\end{minipage}
\caption{Time correlation function of the noise with $\Gamma_3$.}
\label{correlation}
\end{figure}

Figure \ref{noise} shows one colored noise sequence in the case of $\Gamma_3$ with
$W=2$.
Its time correlation function is shown in Fig. \ref{correlation}, where the numerical result
agrees well with the given correlation function. Fifty thousand series of noise have been generated.

We have checked the time evolution of the average kinetic energy, displacement squared,
and the time correlation of the velocity.
The numerical results (not shown) agree well with the analytical ones presented in
the previous subsection,
see Figs. \ref{kernel_G}-\ref{x2}.

\begin{figure}[ht]
\centering
\includegraphics[width = 8cm , height =10cm]{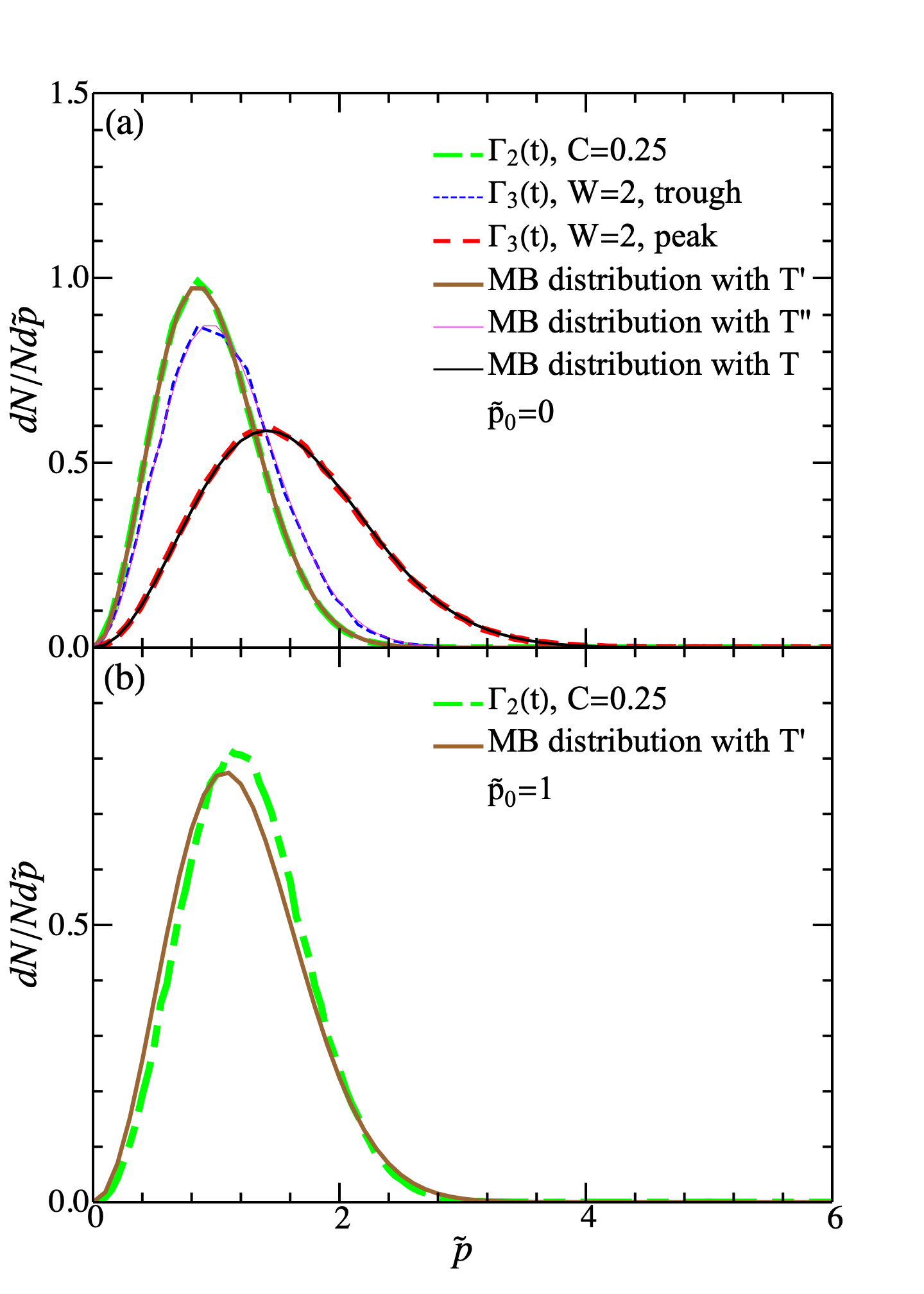}
\caption{The momentum distribution of the Brownian particle in the long-time limit.}
\label{distribution}
\end{figure}

The momentum distributions of an ensemble of Brownian particles at long times are
shown in Fig. \ref{distribution} and compared with the Maxwell-Boltzmann distributions. 
For the case with $\Gamma_3$, the distributions are calculated at the times when the average
kinetic energy reaches its maximum  (peak) as well as its minimum (trough),
see Fig. \ref{kinetic_energy_p0}. From Fig. \ref{distribution} (a), we see that the distribution
of the Brownian particle with $\Gamma_3$ at the peak agrees well with the Maxwell-Boltzmann
distribution with the temperature $T$ of the surrounding matter. This is expected from the study
in the previous subsection that the Brownian particle is in thermal equilibrium at peaks
of the average kinetic energy in the long-time limit. It is also seen, as expected, 
that the momentum distributions of the Brownian particle with $\Gamma_2$ and with 
$\Gamma_3$ at the trough are not in thermal equilibrium with the surrounding matter.
However, they look like the Maxwell-Boltzmann distributions with different ``temperatures''. 
If so, then these ``temperatures'' can be obtained by the values of the average kinetic energy
in the long-time limit according to Eq. (\ref{v2-2}). For the Brownian motion with $\Gamma_2$,
we have
\begin{eqnarray}
\frac{3}{2} k_B T'&=& \frac{1}{2}m<v^2>(t\to \infty) \nonumber \\
&=&\frac{3}{2} k_B T +  \left [  \frac{1}{2}m v^2(0) -  \frac{3}{2} k_B T \right ]
G^2_2(t\to \infty)  \nonumber \\
&=&\frac{3}{2} k_B T +  \left [  \frac{1}{2}m v^2(0) -  \frac{3}{2} k_B T \right ]
\frac{1}{(1+C)^2}  \,,
\label{effectT}
\end{eqnarray}
while for the Brownian motion with $\Gamma_3$ at the trough, we have 
\begin{eqnarray}
\frac{3}{2} k_B T''&=&\frac{3}{2} k_B T +  \left [  \frac{1}{2}m v^2(0) -  \frac{3}{2} k_B T \right ]
\nonumber \\
&& \times G^2_3(t\to \infty, \mbox{ at the trough}) 
\nonumber \\
&=&\frac{3}{2} k_B T +  \left [  \frac{1}{2}m v^2(0) -  \frac{3}{2} k_B T \right ]
\nonumber \\
&& \times \frac{1}{\left (1+\frac{1}{2}\sqrt{\frac{W+1}{W+4}} \right )^2} \,.
\end{eqnarray}
We plot the Maxwell-Boltzmann distributions with $T'$ and $T''$ in Fig. \ref{distribution} (a) and
see agreements of these distributions with the numerical results. 
This is expected by looking at the unidimensional  non-Gaussian indicator \cite{lapas2007entropy}, since
the initial velocity is set to be zero.
The ``temperature'' that the Brownian particle feels is in some cases not
the temperature of the matter, where the Brownian particle is suspended. Therefore,
the attempt to use the Brownian particle to probe an unknown matter may become difficult
because of possible occurrence of anomalous behaviors. 

If the initial condition is nonthermal and the initial velocity is nonzero,
the momentum distribution at a long time is expected to be non-Gaussian, if
$G(t\to \infty)\neq 0$ \cite{lapas2007entropy}. In Fig. \ref{distribution} (b), we show the momentum distribution
of an ensemble of Brownian particles with $\Gamma_2$ and a fixed, nonzero initial
momentum. The distribution is compared with the Maxwell-Boltzmann (Gaussian)
distribution function with a ``temperature'' calculated from Eq. (\ref{effectT}). We see the difference between  two curves, which agrees with the expectation in Ref. \cite{lapas2007entropy}.

\section{The Langevin equation of relativistic Brownian particles}\label{relativity}
Assume that the matter, where the Brownian particle is suspended, remains
at rest. Then, the Langevin equation, Eq. (\ref{langevin2}), can be extended to a form applied
to relativistic particles 
\cite{Debbasch1997RelativisticOP, Debbasch1998ADE, Chevalier2008RelativisticDA, Dunkel:2005zz, Dunkel:2008ngc, Pal_2020, PhysRevE.72.036106},
\begin{equation}
\label{langevin3}
\dot{\boldsymbol{p}}(t)=-\int^{t}_0 dt'  \, \Gamma(t-t') \frac{\boldsymbol{p}(t')c^2}{E(t')} +
\boldsymbol{\xi}(t) \,,
\end{equation} 
where $\boldsymbol{p}$ is the momentum and $E=\sqrt{p^2c^2+m_0^2c^4}$ is the
energy of the Brownian particle with the rest mass $m_0$. Its reduced form with white
noise is
\begin{equation}
\label{langevin4}
\dot{\boldsymbol{p}}=-\gamma \frac{\boldsymbol{p} c^2}{E} + \boldsymbol{\xi} \,.
\end{equation}
The time correlation functions of the noise are same as those in Eqs. (\ref{correl2}) and 
(\ref{correl1}), respectively.

The formal solution of Eq. (\ref{langevin4}) reads
\begin{equation}
\label{solution4}
\boldsymbol{p}(t)=\boldsymbol{p}(t_0)e^{-\gamma  \int_{t_0}^t \frac{dt' c^2}{E}} +
\int_{t_0}^t ds \,  e^{-\gamma \int_s^t \frac{ds' c^2}{E}} \boldsymbol{\xi}(s) \,.
\end{equation}
We then have $<\boldsymbol{p}\cdot \boldsymbol{\xi}>=3\alpha/2$. Taking the ensemble
average of the scalar product of Eq. (\ref{langevin4}) with $\boldsymbol{p}$, we obtain
\begin{equation}
\frac{1}{2}\frac{d}{dt} <p^2>=-\gamma <\frac{p^2c^2}{E}>+\frac{3\alpha}{2} \,.
\end{equation}
If the Brownian particle reaches the thermal equilibrium in the long-time limit, the left-hand
side of the above equation vanishes and $<p^2c^2/E>=3k_B T$ by using the relativistic
Boltzmann distribution $f=\exp[-E/(k_BT)]$. The same fluctuation-dissipation theorem as
Eq. (\ref{dft}) is derived in the relativistic case. Therefore, we assume that
the general fluctuation-dissipation theorem (\ref{dft2}) is also valid for relativistic Brownian
particles because a memory kernel having the $\delta$-function form reduces the Langevin 
equation from Eq. (\ref{langevin3}) to Eq. (\ref{langevin4}) and the fluctuation-dissipation
theorem from Eq. (\ref{dft2}) to Eq. (\ref{dft}). We notice that from the assumption of the
fluctuation-dissipation theorem, we can also derive that the relativistic Brownian particle
under white noise will reach the thermal equilibrium in the long-time limit.

It is obvious that both Eqs. (\ref{langevin3}) and (\ref{langevin4}) are not linear
differential-integral equations and thus cannot be solved analytically by using
the Laplace transformation as performed in the nonrelativistic case. Even the solution
(\ref{solution4}) can only be calculated numerically, since $E$ on the right-hand side of
Eq. (\ref{solution4}) is a function of $p$ at times before $t$. Even though we can derive that
the Brownian particle under white noise can reach the thermal equilibrium in the long-time
limit, we cannot determine the behavior of the diffusion and ergodicity analytically. 

In the following, we solve the Langevin equations, Eqs. (\ref{langevin3}) and (\ref{langevin4}),
by using the numerical code mentioned and well tested in the previous section. 
The noise is approximately Gaussian \cite{PhysRevE.74.051106}. We will study
the equilibration, memory effect, diffusion, and ergodicity of Brownian particles under
white noise and colored noise with the memory kernels $\Gamma_1$, $\Gamma_2$,
and $\Gamma_3$ given in the previous section.
 
Like Eq. (\ref{langevin_nond}), the relativistic Langevin equation, Eq. (\ref{langevin3}), can
also be nondimensionalized, when we define $\tilde{\boldsymbol{p}}=\boldsymbol{p}c/(k_B T)$,
$\tilde{m}_0=m_0 c^2/(k_B T)$, $\tilde{t}=t/\tau$, and 
$\tilde{\boldsymbol{\xi}}=\boldsymbol{\xi} \tau c/(k_B T)$. We have then
\begin{equation}
\label{langevin_nond2}
\frac{d\tilde{\boldsymbol{p}}}{d\tilde{t}}=-\int^{\tilde{t}}_0 d\tilde{t}' \, 
\tilde{\Gamma}(\tilde{t}-\tilde{t}') 
\frac{\tilde{\boldsymbol{p}}(\tilde{t}')}{\sqrt{\tilde{p}^2+\tilde{m}_0^2}}+
\tilde{\boldsymbol{\xi}}(\tilde{t})\,,
\end{equation}
where
\begin{equation}
\tilde{\Gamma}(\tilde{t}-\tilde{t}')= \frac{\tau^2c^2}{k_B T} \Gamma(t-t')
\end{equation}
and
\begin{equation}
<\tilde{\xi}_i(\tilde{t}) \tilde{\xi}_j(\tilde{t}')>=\delta_{ij} \tilde{\Gamma}(\tilde{t}-\tilde{t}')\,.
\end{equation}
$\tau$ is the decay timescale in the given memory kernels. Looking at the memory kernels
$\Gamma_1$, $\Gamma_2$, and $\Gamma_3$ in Eqs. (\ref{gm1}), (\ref{gm2}), and 
(\ref{gm3}), we find that $\tilde{m}_0$ and $\gamma \tau c^2/(k_B T)$
are the free parameters in Eq. (\ref{langevin_nond2}). 
Note that $\omega \tau$ in $\Gamma_3$ is not an additional parameter, since
$\omega \tau$ (or $W$) is fixed by $\gamma\tau/m_0$ in Eq. (\ref{parameter}), where
we replace $m$ by $m_0$. We also realize that $\tau$ does not exist in the Langevin 
equation with white noise, Eq. (\ref{langevin4}). In this case, we define
$\tau=k_B T/(\gamma c^2)$.

\begin{figure}[ht]
\centering
\includegraphics[width = 8cm , height =7cm]{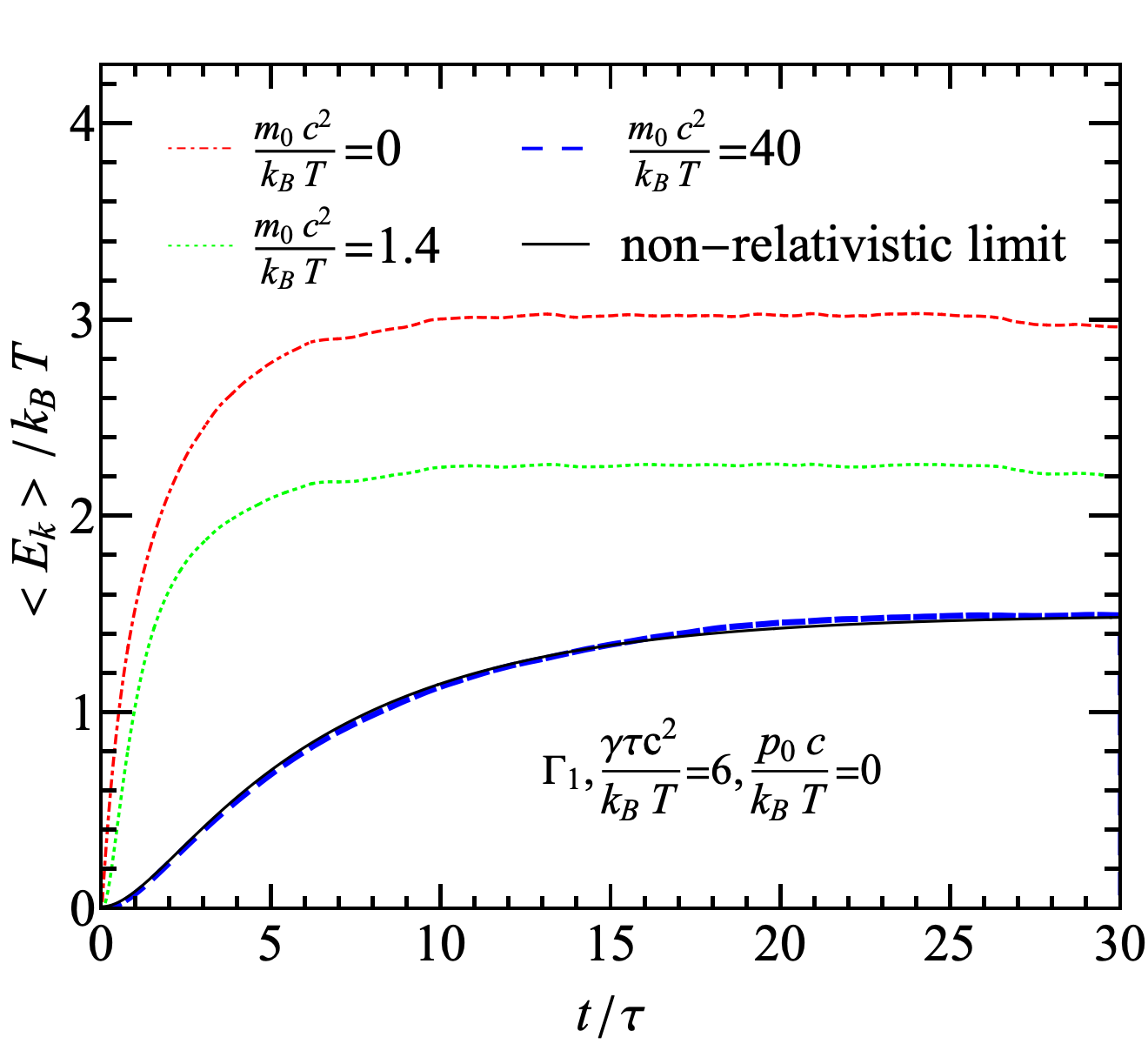}
\caption{The time evolution of the average kinetic energy of the relativistic Brownian particle
under colored noise with $\Gamma_1$ and with various rest masses. The initial momentum
of the particle is set to be zero.}
\label{relmass}
\end{figure}

Figure \ref{relmass} shows the time evolution of the average kinetic energy of the relativistic
Brownian particle, $E_k=E-m_0c^2$ scaled by $k_B T$, under colored noise with
$\Gamma_1$. 
The initial momentum of the Brownian particle is zero. The parameter
$\gamma \tau c^2/(k_B T)$ is set to be $6$. For $\tau=1 \text{ fm/c}$ and $T=300 \text{ MeV}$,
$\gamma/2$ is consistent with the drag coefficient in the Brownian motion
of charm quarks in the quark-gluon plasma created in relativistic heavy-ion collisions
\cite{Cao:2015cba, Dong:2019unq, Akiba:2015jwa, vanHees:2005wb, PHENIX:2006iih, ALICE:2020iug, Banerjee:2011ra, PhysRevD.86.014509, Moore:2004tg}.  
We vary the rest mass of the Brownian particle to see the relativistic effect and
to check the numerical calculations.

We see that in the long-time limit, the average kinetic energy increases from the nonrelativistic
limit, $3k_BT/2$, to the ultrarelativistic limit, $3k_B T$. All the final values with different
masses agree with those according to the equipartition theorem, which indicates that the
Brownian particle under colored noise with $\Gamma_1$ reaches the thermal equilibrium
with its surrounding matter. The numerical result of the relativistic Langevin equation with a 
large mass $\tilde{m}_0=40$ is compared with the analytical one of the nonrelativistic Langevin
equation. The perfect agreement demonstrates the solid numerical calculations.

\begin{figure}[h]
\centering
\includegraphics[width = 8cm , height =7cm]{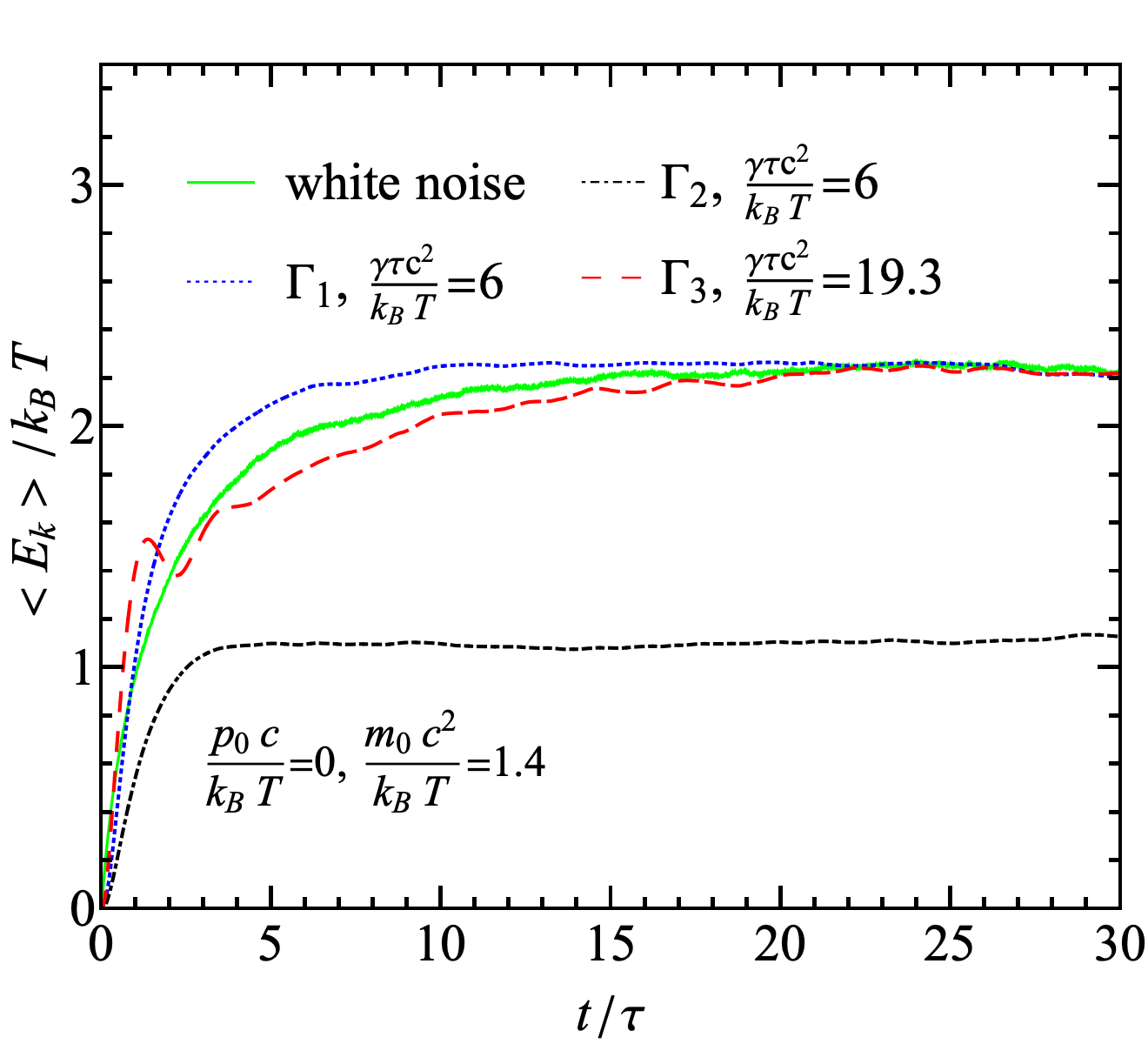}
\caption{Same as Fig. \ref{relmass} but under white noise and colored noise with
various memory kernels.}
\label{re_kinetic_energy_p0}
\end{figure}

The time evolution of the average kinetic energy of the relativistic Brownian particle
with mass $\tilde{m}_0=1.4$ under white noise and colored noise 
with the memory kernels $\Gamma_1$, $\Gamma_2$, and $\Gamma_3$ are depicted
in Fig. \ref{re_kinetic_energy_p0}. 
The initial momentum is zero. The curve of $\Gamma_1$ is the same as that in 
Fig. \ref{relmass}. For $\Gamma_3$ with $W=2$, we obtain $\gamma \tau/m_0=13.8$
according to Eq. (\ref{parameter}) by replacing $m$ by $m_0$. Thus, we have
$\gamma \tau c^2/(k_B T)=19.3$ for $\tilde{m}_0=1.4$.

From Fig. \ref{re_kinetic_energy_p0}, we see that the average kinetic energy of
the Brownian particle under white noise and colored noise with $\Gamma_1$ and
$\Gamma_3$ approaches the same value in the long-time limit, which is the value
according to the equipartition theorem. The oscillating behavior in the nonrelativistic limit with 
$\Gamma_3$ does not occur here. Remember that Eq. (\ref{parameter}) is a strict condition for
the oscillating behavior in the nonrelativistic case, which is not necessarily the same one met
in the relativistic case.

The long-time limit of the average kinetic energy of the Brownian particle with $\Gamma_2$ is
smaller than that with other memory kernels. The anomalous behavior already seen in the
nonrelativistic case appears in the relativistic case too. We want to know whether we can
make use of the formulas derived in the nonrelativistic limit in the previous section to analyze
the results achieved in the relativistic case. From Eq. (\ref{v2-2}), we get the average kinetic
energy in the long-time limit,
\begin{eqnarray}
\label{approx}
\frac{<E_k>|_{t\to \infty}}{k_B T}&=&\frac{<E_k>|_{t=0}}{k_B T}\frac{1}{(1+C)^2}\nonumber \\
&&+\frac{3}{2} \left [ 1-\frac{1}{(1+C)^2} \right ] \,,
\end{eqnarray}
where $E_k=p^2/(2m)$ and $C=\gamma \tau/(4m)$. We use this formula and replace 
$m$ by $m_0$. For $p_0=0$, $\gamma \tau c^2/(k_B T)=6$, and $\tilde{m}_0=1.4$, 
we obtain $<E_k>|_{t\to \infty}/(k_B T)=1.15$, which agrees
with the numerical result seen in Fig. \ref{re_kinetic_energy_p0}.

\begin{figure}[h]
\centering
\includegraphics[width = 8cm , height =7cm]{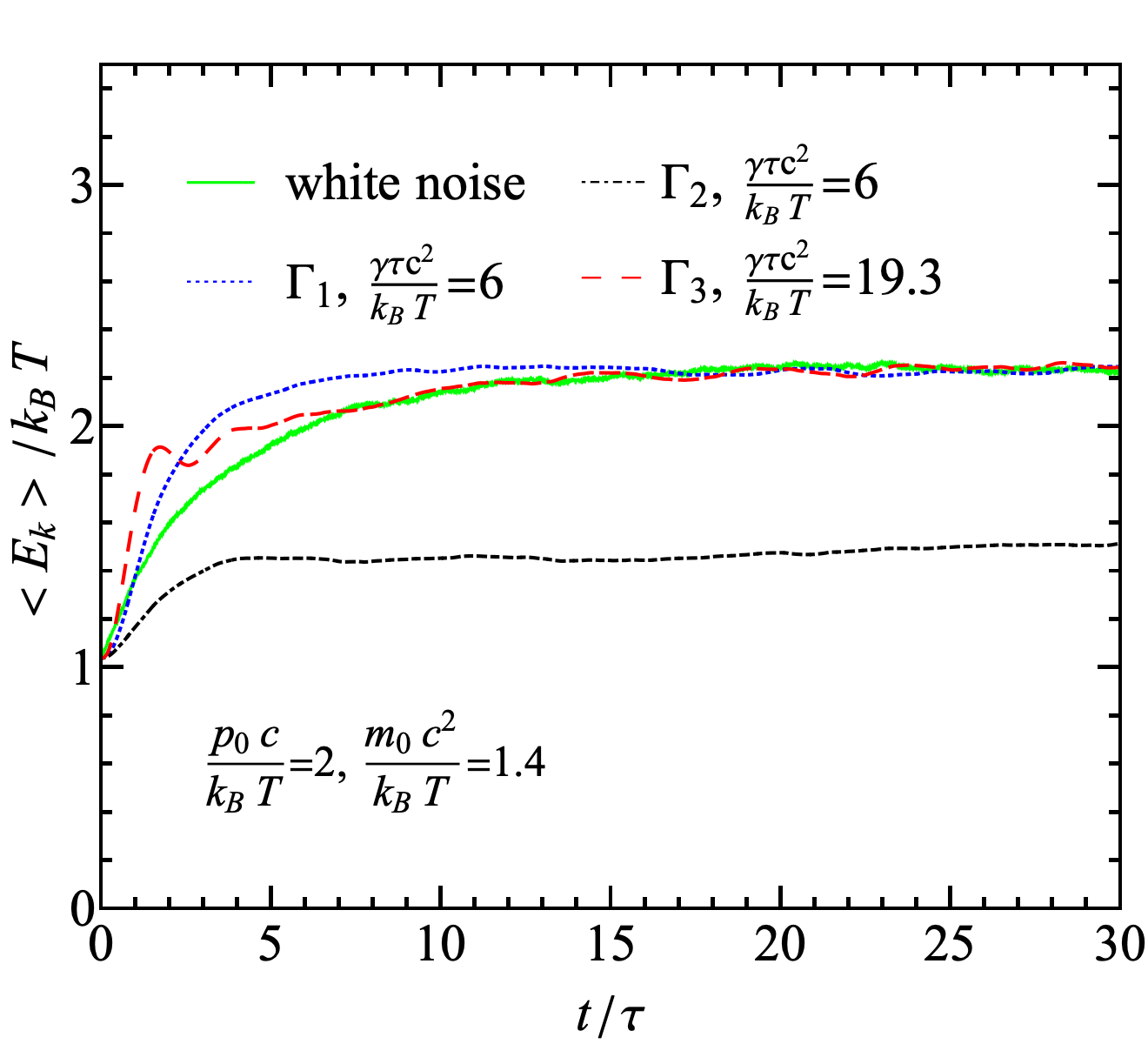}
\caption{Same as Fig. \ref{re_kinetic_energy_p0}, but the initial momentum of the particle
is set to be $p_0c/(k_B T)=2$.}
\label{re_kinetic_energy_p2}
\end{figure}

The memory effect of the initial momentum is shown in Fig. \ref{re_kinetic_energy_p2}.
In these calculations, the initial momentum is set to be $p_0c/(k_B T)=2$. By comparing
with Fig. \ref{re_kinetic_energy_p0}, we see that
while the long-time limit of the average kinetic energy of the relativistic Brownian particle
under white noise and colored noise with $\Gamma_1$ and $\Gamma_3$ do not
show any dependence of the initial momentum, it does appear for the particle with
$\Gamma_2$. Using the formula in the nonrelativistic limit (\ref{approx}), we obtain 
$<E_k>|_{t\to \infty}/(k_B T)=1.48$ for $p_0c/(k_B T)=2$, which also agrees
with the numerical result seen in Fig. \ref{re_kinetic_energy_p2}.

\begin{figure}[t]
\centering
\includegraphics[width = 8cm , height =7cm]{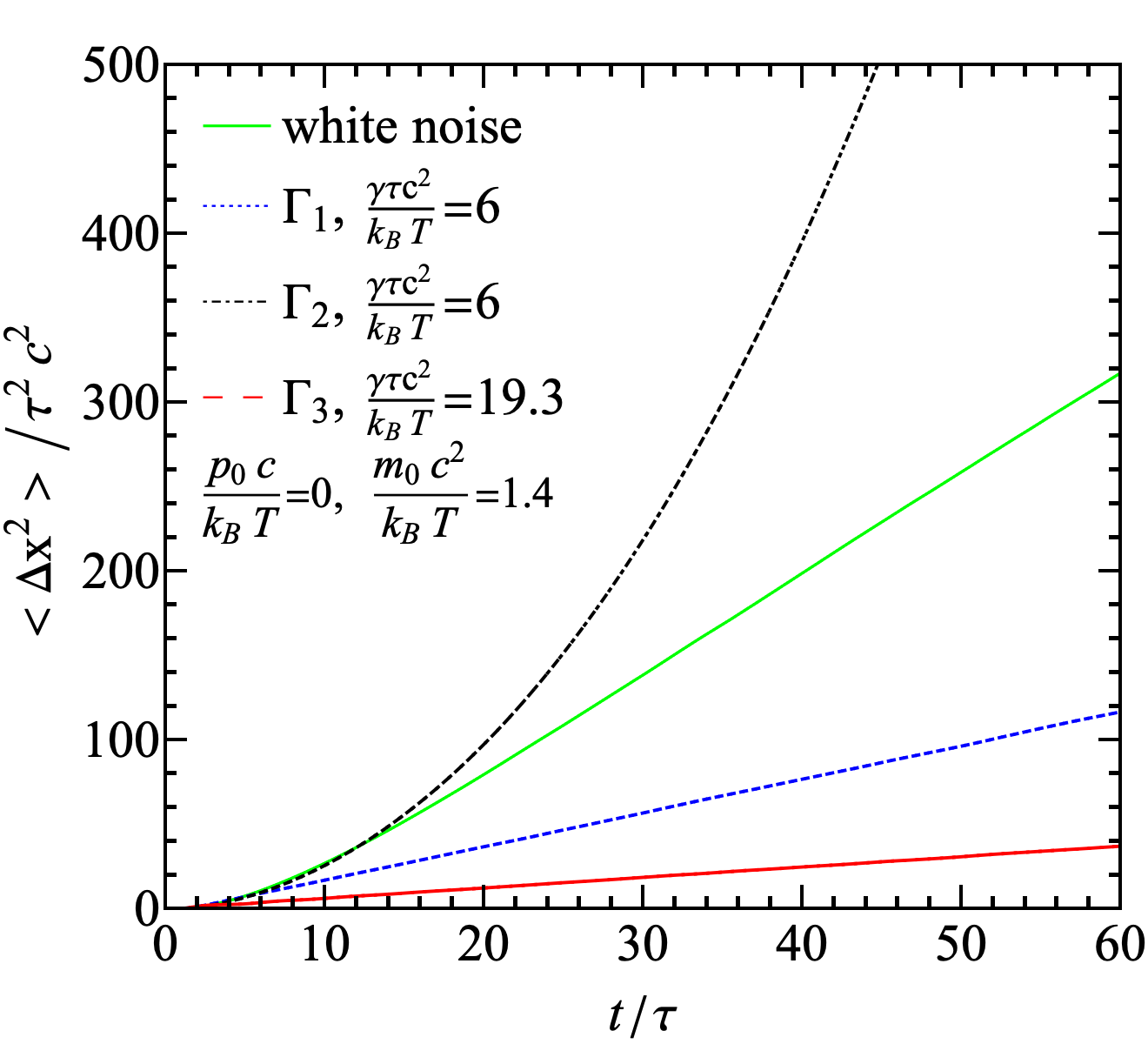}
\caption{The time evolution of the average displacement squared of the relativistic Brownian
particle under white noise and colored noise with various memory kernels. The initial
momentum of the particle is set to be $p_0=0$. }
\label{re_x2}
\end{figure}

We depict the time evolution of the average displacement squared, scaled by $\tau^2 c^2$,
in Fig. \ref{re_x2}.
The behaviors of the diffusion are the same as in the nonrelativistic limit. The particle diffusion
under white noise and colored noise with $\Gamma_1$ and $\Gamma_3$ are normal,
while it is ballistic for the particle with $\Gamma_2$. Using the formulas in the nonrelativistic
limit Eqs. (\ref{difflimit_wn})-(\ref{difflimit_3}) times $k_B T/(m_0 c^2)$,
\begin{eqnarray}
\label{re_difflimit_wn}
&&\frac{<(\Delta \boldsymbol{x})^2>|_{\mbox{wn}}}{\tau^2 c^2} \to 
6 \frac{k_B T}{m_0 c^2} \frac{t}{\tau} \,, \\
\label{re_difflimit_1} 
&&\frac{<(\Delta \boldsymbol{x})^2>|_{\Gamma_1, \Gamma_3}}{\tau^2 c^2} \to 
12 \frac{k_B T}{\gamma \tau c^2} \frac{t}{\tau} \,, \\
\label{re_difflimit_2}
&& \frac{<(\Delta \boldsymbol{x})^2>|_{\Gamma_2}}{\tau^2 c^2} \to
\frac{k_B T}{m_0 c^2}
 \frac{p_0^2/(m_0 k_B T)+3C}{(1+C)^2} \left ( \frac{t}{\tau} \right )^2\,,
\end{eqnarray}
we compare these nonrelativistic results with the relativistic ones. We find perfect
agreements for colored noise with $\Gamma_1$ and $\Gamma_3$, an approximate 
agreement for white noise, and a difference of almost a factor of $2$ for colored
noise with $\Gamma_2$.

Finally, we calculate the autocorrelation function of momentum at different times to
study the validity of the ergodicity in the motion of relativistic Brownian particles. 
Figure 12 shows the numerical results for the initial momentum $\tilde{p}_0=1$
with a random direction.
\begin{figure}[ht]
\centering
\includegraphics[width = 8cm , height =7cm]{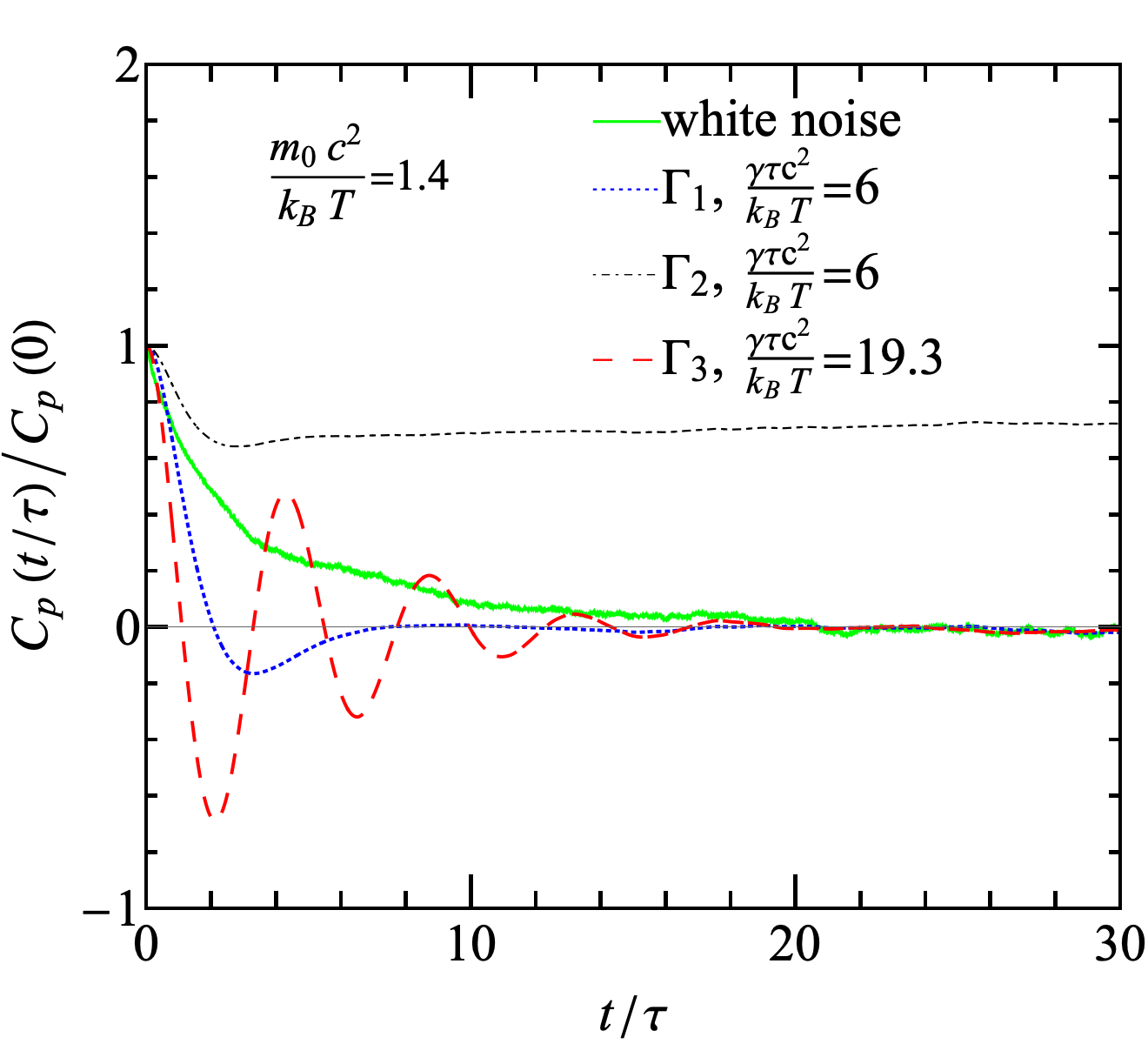}
\caption{The autocorrelation function of momentum of the relativistic Brownian particle under
white noise and colored noise with various memory kernels.}
\label{re_corr_p0}
\end{figure}
The momentum is scaled by $k_B T/c$. We clearly see that $C_p$ with $\Gamma_2$ is
nonzero in the long-time limit, while the other three correlations approach zero.
Therefore, the motion with $\Gamma_2$ breaks the validity of the ergodicity, while
the motions under white noise and colored noise with $\Gamma_1$ and 
$\Gamma_3$ hold the ergodicity.

\section{Summary}
\label{sum}
In this paper, we solved analytically the generalized Langevin equation of nonrelativistic
Brownian particles with the memory kernel and colored noise by employing the Laplace
transformation technique. According to the position of the singularities of the response function 
$G(s)$, the memory kernels are classified to four categories, which result in different 
behaviors of the thermal equilibrium, the memory effect, the particle diffusion, and the
ergodicity. Specifically, the first category includes the cases that all the poles of $G(s)$ are
located in the left half of the complex plane but not on the imaginary axis. In the long-time limit,
the Brownian particle approaches thermal equilibrium with the surrounding matter,
has no memory of the initial state, and diffuses normally. The ergodicity holds. 
The second category includes the cases that one pole is located on the imaginal axis
and specially at the origin $s=0$ and the other poles are located in the left half of the complex
plane. Usually, the Brownian particle cannot reach the equilibrium
with the surrounding matter, but will reach an equilibrium with a different temperature
rather than that of the surrounding matter, if the initial velocity is zero. 
This temperature as well as the average kinetic
energy of the Brownian particle depend on its initial state, which shows the memory effect.
The particle diffusion is ballistic and the ergodicity is broken.
The third category includes the cases that some poles in pairs are located on the
imaginal axis but not at the origin and the other poles are located in the left half of the complex
plane. The Brownian particle approaches and departs from the thermal equilibrium
with the surrounding matter periodically. The amplitude of the oscillation in the average
kinetic energy depends on the initial state, which is again the memory effect. The diffusion
of the Brownian particle is normal, but the ergodicity is broken.
The fourth and last category includes the memory kernels with the form 
$\lim_{s\to 0} \Gamma(s)=\gamma(s\tau)^{\lambda-1}$, which has a branch point at $s=0$.
For $0<\lambda<2$, $G(t\to \infty)$ goes to zero. The long-time behaviors of the Brownian
particle are the following. The particle approaches the thermal equilibrium with the
surrounding matter and has no memory of the initial state. For $0<\lambda<1$, the diffusion
is the subdiffusion, while it is superdiffusion for $1<\lambda<2$. The ergodicity holds. For
$\lambda >2$, the branch point is replaced by the pole at $s=0$ in the long-time limit.
The behaviors of the Brownian particle are the same as those in the second category.

For relativistic Brownian particles, the Langevin equation cannot be solved analytically 
because of the nonlinearity. Therefore, we do not have a general conclusion about
the behavior of the thermal equilibrium, memory effects, diffusion, and ergodicity
for any given memory kernel. On the other hand, we solved the relativistic Langevin
equation numerically for three typical memory kernels chosen as the examples in 
the nonrelativistic case. Similar results as obtained in the nonrelativistic case for the first
and second categories of memory kernels are also seen in the relativistic case. In other
words, there are indeed memory kernels, with which the relativistic Brownian particle cannot
reach thermal equilibrium, has memory effects of the initial state, and diffuses anomalously.
Its motion breaks the ergodicty. Moreover, by regarding the relativistic particle as the
nonrelativistic one (by replacing the mass by the rest mass), the average kinetic energy and
the average displacement squared (except for one case with $\Gamma_2$) can be well
described by the formulas derived in the nonrelativistic case.

Persistent nonequilibrium effects in Brownian motions may challenge the probe of 
an unknown matter
by using a Brownian particle. On the other hand, our present investigation may give rise to
think about the memory effect and anomalous diffusion of heavy quarks in the quark-gluon
plasma created in relativistic heavy-ion collisions.

\acknowledgments
This work was financially supported by the National Natural Science Foundation of China
under Grants No. 11890710, No. 11890712, and No. 12035006, the Ministry of Science and
Technology under Grant No. 2020YFE0202001.
C.G. acknowledges support by the Deutsche Forschungsgemeinschaft (DFG) through
the grant CRC-TR 211 ``Strong-interaction matter under extreme conditions.''

%%%%%%%%%%%%%%   Appendix    %%%%%%%%%%%%%%
\appendix 
\section{Laplace transformation}
\label{app1}
The Laplace transform of a function $f(t)$ is defined as
\begin{equation}
L[f(t)]=\int_0^t dt f(t) e^{-st}\equiv f(s) \,,
\end{equation}
where $s$ is a complex variable. The inverse transform is 
\begin{equation}
L^{-1}[f(s)]=\frac{1}{2\pi i} \int_{\beta -i\infty}^{\beta+i\infty} ds f(s) e^{st} \,.
\end{equation} 
Some useful properties of the Laplace transformation are listed below:
\begin{eqnarray}
&&L\left [ \frac{df(t)}{dt} \right ]=sf(s)-f(t=0) \,,\\
&&L\left [\int_0^t dt' f(t-t') g(t') \right ] = f(s)g(s)\,, \\
&&L\left [ \frac{d^2f(t)}{dt^2} \right ]=s^2f(s)-sf(t=0)-\frac{df(t)}{dt}(t=0) \,.
\end{eqnarray}

We now derive Eq. (\ref{v2-2}) from Eq. (\ref{v2-1})
\begin{eqnarray}
<v^2>(t)&=&v^2(0)G^2(t)+\frac{3k_BT}{m^2}\int_0^t dt' G(t-t')   \nonumber \\
&& \times \int_0^t dt'' G(t-t'') \Gamma(t''-t') \nonumber \,.
\end{eqnarray}
Because the integrals for $t'' \ge t'$ and $t'' \le t'$ are same, we rewrite the above equation to
\begin{eqnarray}
\label{appv2-1}
<v^2>(t)&=&v^2(0)G^2(t)+\frac{3k_BT}{m^2}2 \int_0^t dt' G(t-t')  \nonumber \\
&& \times \int_0^t dt'' G(t-t'') \Gamma(t''-t') \theta(t''-t')  \,.
\end{eqnarray}
By $\tau=t''-t'$, we have
\begin{eqnarray}
&&\int_0^t dt'' G(t-t'') \Gamma(t''-t') \theta(t''-t') \nonumber \\
&=&\int_{-t'}^{t-t'} d\tau G(t-t'-\tau) \Gamma(\tau) \theta(\tau) \nonumber \\
&=& \int_0^{t-t'} d\tau G(t-t'-\tau) \Gamma(\tau) \nonumber \\
&=& L^{-1}\left \{ L \left [ \int_0^{t-t'} d\tau G(t-t'-\tau) \Gamma(\tau) \right ] \right \} \nonumber \\
&=& L^{-1} \left [ G(s) \Gamma(s) \right ] \nonumber \,. 
\end{eqnarray}
Since $G(s)=1/[s+\Gamma(s)/m]$ and $G(t=0)=1$, we get 
\begin{equation}
G(s)\Gamma(s)=m [sG(s)-G(t=0)]=m L\left [ \frac{d G(t-t')}{d(t-t')} \right ] \nonumber
\end{equation}
and
\begin{equation}
L^{-1} \left [ G(s)\Gamma(s) \right ] =m \frac{d G(t-t')}{d(t-t')} 
= -m \frac{d G(t-t')}{dt'}  \nonumber \,.
\end{equation}
Putting this in Eq. (\ref{appv2-1}), we finally obtain
\begin{eqnarray}
<v^2>(t)&=&v^2(0)G^2(t)-\frac{3k_BT}{m}2 \int_0^t dt' G(t-t') \frac{d G(t-t')}{dt'}  \nonumber \\
&=&v^2(0) G^2(t) + \frac{3k_B T}{m}\left [1 - G^2(t) \right ] \nonumber \,.
\end{eqnarray}

%%%%%%%%%%%%   References   %%%%%%%%%%%%%%%%
\bibliographystyle{apsrev}
%\bibliography{nucleation}{}
%\bibliographystyle{apsrev4-1}
%\bibliography{References}{}

\end{document}